\documentclass[fleqn,usenatbib]{mnras}

\usepackage{newtxtext,newtxmath}

\usepackage[T1]{fontenc}

\DeclareRobustCommand{\VAN}[3]{#2}
\let\VANthebibliography\thebibliography
\def\thebibliography{\DeclareRobustCommand{\VAN}[3]{##3}\VANthebibliography}

\usepackage{graphicx}	
\usepackage{amsmath}	

\newcommand{\msun}{M$_{\odot}$}
\newcommand{\zsun}{Z$_{\odot}$}
\newcommand{\enzo}{\texttt{Enzo}}

\title[Why does the Milky Way have a metallicity floor?]{Why does the Milky Way have a metallicity floor?}

\author[B. D. Smith et al.]{
Britton D. Smith,$^{1}$\thanks{E-mail: brs@roe.ac.uk (BDS)}
Brian W. O'Shea,$^{2,3,4}$
Sadegh Khochfar,$^{1}$
Matthew J. Turk,$^{5,6}$
John H. Wise,$^{7}$
\newauthor
and Michael L. Norman$^{8,9}$
\\
$^{1}$Institute for Astronomy, University of Edinburgh, Royal Observatory, Edinburgh EH9 3HJ, UK\\
$^{2}$Department of Computational Mathematics, Science, and Engineering, Michigan State University, East Lansing, MI 48824, USA\\
$^{3}$Department of Physics and Astronomy, Michigan State University, East Lansing, MI 48824, USA\\
$^{4}$Facility for Rare Isotope Beams, Michigan State University, East Lansing, MI 48824, USA\\
$^{5}$School of Information Sciences, University of Illinois, Urbana-Champaign, IL, 61820, USA\\
$^{6}$Department of Astronomy, University of Illinois, Urbana-Champaign, IL, 61820, USA\\
$^{7}$Center for Relativistic Astrophysics, School of Physics, Georgia Institute of Technology, Atlanta, GA, 30332, USA\\
$^{8}$San Diego Supercomputer Center, University of California, San Diego, 10100 Hopkins Drive, La Jolla, CA 92093, USA\\
$^{9}$Center for Astrophysics and Space Sciences, University of California, San Diego, 9500 Gilman Dr, La Jolla, CA 92093, USA
}

\date{Accepted XXX. Received YYY; in original form ZZZ}

\pubyear{2024}

\begin{document}
\label{firstpage}
\pagerange{\pageref{firstpage}--\pageref{lastpage}}
\maketitle

\begin{abstract}
The prevalence of light element enhancement in the most metal-poor stars is potentially an indication that the Milky Way has a metallicity floor for star formation around $\sim$10$^{-3.5}$ \zsun. We propose that this metallicity floor has its origins in metal-enriched star formation in the minihalos present during the Galaxy's initial formation. To arrive at this conclusion, we analyze a cosmological radiation hydrodynamics simulation that follows the concurrent evolution of multiple Population III star-forming minihalos. The main driver for the central gas within minihalos is the steady increase in hydrostatic pressure as the halos grow. We incorporate this insight into a hybrid one-zone model that switches between pressure-confined and modified free-fall modes to evolve the gas density with time according to the ratio of the free-fall and sound-crossing timescales. This model is able to accurately reproduce the density and chemo-thermal evolution of the gas in each of the simulated minihalos up to the point of runaway collapse. We then use this model to investigate how the gas responds to the absence of H$_{2}$. Without metals, the central gas becomes increasingly stable against collapse as it grows to the atomic cooling limit. When metals are present in the halo at a level of $\sim$10$^{-3.7}$ \zsun, however, the gas is able to achieve gravitational instability while still in the minihalo regime. Thus, we conclude that the Galaxy's metallicity floor is set by the balance within minihalos of gas-phase metal cooling and the radiation background associated with its early formation environment.
\end{abstract}




\section{Introduction}

The abundance patterns of extremely metal-pool stars comprise nearly all of the observational data that we have on the process of star formation below a metallicity of about 10$^{-3}$ \zsun. One of the most notable features of these objects, collectively, is their enhancement in light elements, namely C and O, with respect to Fe, the classical observational proxy for ``metallicity\footnote{While iron abundance, [Fe/H], has often been used historically to denote metallicity, we make a distinction in this work between individual elemental abundances and metallicity. We use metallicity, $Z$, to mean the fraction of total gas \textit{mass} in metals, whereas elemental abundances are typically expressed as the ratio of two elements, X and Y, by [X/Y] $\equiv \log_{10}(N_X/N_Y) - \log_{10}(N_X/N_Y)_{\odot}$, where $N_X$ is the \textit{number} abundance of element X.}'' From this, we have the notion of Carbon-Enhanced Metal-Poor (CEMP) stars \citep{2005ARA&A..43..531B}, which comprise a majority of stars with [Fe/H] $<$ -3. If we consider the combined metal content of these stars, however, another interpretation of the CEMP phenomenon is that of a metallicity floor \citep[e.g.,][]{2018MNRAS.481.3838S}. Indeed, as noted there and elsewhere, [C/Fe] seems to increase with decreasing [Fe/H] for the most metal-poor objects. This point was made succinctly by \citet{2007MNRAS.380L..40F}, who introduce a combined C/O abundance measure known as the transition discriminant ($D_{\rm trans}$), given by
\begin{equation}
    \label{eqn:dtrans}
    D_{\rm trans} \equiv \mathrm{log}_{10} (10^{\mathrm{[C/H]}} + 0.3 \times 10^{\mathrm{[O/H]}}).
\end{equation}
This particular abundance measure is designed to highlight the most important coolants in low metallicity gas (see below and Section \ref{sec:interpretation}). \citet{2007MNRAS.380L..40F} find a minimum value for stars in the Milky Way of $D_{trans} \sim -3.5$ (Figure 1 of that work), equivalent to an absolute metallicity of about 10$^{-3.6}$ \zsun. Thus, when viewed from this perspective, the Galaxy appears to have a metallicity floor.

This number has very interesting astrophysical significance. Well above this metallicity, observations overwhelmingly support the existence of a roughly universal stellar initial mass function (IMF), robust to physical conditions and producing predominantly low mass stars \citep[see, e.g.,][]{2014PhR...539...49K}\footnote{In practice, direct observational evidence of the IMF extends down to around $\sim$10$^{-2}$ \zsun\ \citep[see the discussion in][]{2022MNRAS.509.1959S}.}. Well below it (specifically, zero metallicity), we expect the star formation process to have fundamentally different results, i.e., a distinctly top-heavy IMF that may also have environmental dependence \citep[see, e.g.,][]{2023ARA&A..61...65K}. To zeroth order we know that H$_{2}$ (the main coolant in cold, metal-free gas) plants the initial seed for a top-heavy IMF through its inefficient cooling, which limits Jeans fragmentation to mass scales of $\sim$1000 \msun\footnote{We note that most simulations now show the dense accretion disk surrounding the massive pre-stellar core will readily fragment into small clumps \citep[again, see][]{2023ARA&A..61...65K}, but this does not alter the main conclusion.}. If we then consider the point during the collapse of metal-free gas where H$_{2}$ cooling becomes inefficient and fragmentation ends, it is possible to calculate the amount of metal required for continued cooling and fragmentation, arriving at a metallicity of $\sim$10$^{-3.5}$ \zsun. Specifically, this is done by equating the cooling time with the free-fall time at the relevant density and temperature ($n \sim 10^{4}$ cm$^{-3}$ and $T \sim 200$ K). What's more, as was first pointed out by \citet{2003Natur.425..812B}, the most important metal coolants in this regime come from atomic C and O. Thus, it is tempting to conclude that the Galaxy's metallicity floor traces the ``gas-phase critical metallicity'' for the transition from Population (Pop) III to Pop II star formation. 

Problems exist with this interpretation, however. Another significantly lower critical metallicity for fragmentation exists due to cooling from dust \citep{2005ApJ...626..627O}, with a value of $\sim$10$^{-5.5}$ \zsun\ \citep[][assuming a local ISM dust-to-metal ratio]{2015MNRAS.446.2659C}. Decoupling dust from gas-phase metallicity, this value has also been expressed as a critical ``dusticity,'' $D_{cr} \sim 5\times10^{-9}$ \citep{2012MNRAS.419.1566S}\footnote{This is about $\sim$5$\times10^{-7}$ of the local dust-to-gas ratio of 0.009387 \citep{1994ApJ...421..615P}.}, still far less total metal than the previously stated gas-phase critical metallicity. As well, both observations and simulations indicate that the Universe has the ability to create low mass stars above the dust-phase critical metallicity but below the gas-phase one. Famously, the star SDSS J102915+172927 \citep{2011Natur.477...67C} has [Fe/H] $\sim$ -5 and [C/H] $\sim$ -4.4. Complementing that, the fully cosmological simulations of \citet{2015MNRAS.452.2822S} showed that such a low metallicity (in that work, $Z \sim 2\times10^{-5}$ \zsun) star-forming environment could arise in a minihalo on the cusp of Pop III star formation that is enriched by a supernova blast-wave originating from another nearby minihalo, i.e., prompt external enrichment. In summary, the Galactic metallicity floor does not appear to be a signal of IMF transition.

It seems logical, though, that the physical context in which to place the supposed metallicity floor is enriched minihalos. Cosmological simulations show them to be particularly capable of forming extremely metal-poor stars, either by external \citep{2021ApJ...909...70H} or internal enrichment \citep[i.e., fallback of supernova ejecta][]{2018MNRAS.475.4378C}. Crucially, we define a minihalo as a dark matter halo with virial temperature too low to cool via Lyman-$\alpha$ emission (i.e., less than $\sim$10$^{4}$ K). Thus, without metals it must cool via H$_{2}$. Once again, it becomes tempting to try and equate H$_{2}$ and metal cooling rates. And so we have. In this paper, we present a novel explanation for the existence of a Galactic metallicity floor in which minihalos must reach a minimum metallicity to form stars in the presence of a photodissociating Lyman-Werner background associated with the early formation environment of the Milky Way. We arrive at this result via a somewhat indirect route. First, we use an extremely high resolution cosmological simulation to characterize the origin of gravitationally unstable gas in minihalos. We then utilize the lessons learned to construct a one-zone model capable of reproducing the long term evolution of minihalo gas. Following this, we employ the model to investigate the reaction of these halos to the loss of H$_{2}$ and the metals required to supplement this. Finally, we introduce our interpretation of the metallicity floor and/or CEMP stars. In Section \ref{sec:simulation} we describe the simulation that forms the foundation of this work. In Section \ref{sec:minihalos} we present the relevant results from the simulation, the subsequent ``Minihalo model,'' and the most important metallicity thought experiment. We discuss our key interpretation in Section \ref{sec:interpretation}, potential caveats in Section \ref{sec:caveats}, and provide a summary of results and concluding remarks in Section \ref{sec:conclusion}.

\section{The Simulation}
\label{sec:simulation}

The simulation used in this work is of the \texttt{Pop2Prime} family of simulations. It is a variant of the one presented in \citet{2015MNRAS.452.2822S}, with two key differences. The simulation of the previous work was designed to investigate the conditions leading to metal-enriched star formation above the dust-phase critical metallicity, whereas here the goal is to do the same, but for the gas-phase critical metallicity. Hence, the two changes are an increase in the threshold metallicity for Pop III star formation from 10$^{-6}$ \zsun\ to 10$^{-4}$ \zsun\ and the exclusion of dust grains in the chemistry and cooling. We briefly describe the simulation code and setup below but refer the reader to that work for a thorough discussion of the methodology. We also refer the reader to \citet{2024MNRAS.527..307C} for a description of the evolution of the Pop III star forming minihalos in the simulation.

\subsection{Simulation Code}
\label{sec:enzo}

We use the \enzo\ simulation code \citep{2014ApJS..211...19B, 2019JOSS....4.1636B} for the work presented here. \enzo\ is an open source, adaptive mesh refinement (AMR) + N-body cosmological simulation code that has been used to study a wide variety of astrophysical phenomena, especially in the era of Cosmic Dawn. \enzo\ solves the equations of ideal hydrodynamics (adapted for cosmology) in the comoving frame using a block-structured Eulerian AMR framework \citep{1989JCoPh..82...64B} to dynamically add resolution to regions of interest based on several criteria. Here we employ the Piecewise Parabolic Method hydro solver of \citet{1984JCoPh..54..174C} coupled to an N-body adaptive particle-mesh gravity solver \citep{1985ApJS...57..241E, 1988csup.book.....H}. We use the \texttt{Moray} adaptive ray-tracing radiation transport solver \citep{2011MNRAS.414.3458W} to propagate H/He ionizing radiation from individual Pop III stars. We model the chemistry and radiative cooling of the gas using machinery that was, at the time of simulation, a precursor to the \texttt{Grackle} library \citep{2017MNRAS.466.2217S}. This solves a non-equilibrium network of 12 primordial species (H, H$^{+}$, He, He$^{+}$, He$^{++}$, e$^{-}$, H$_{2}$, H$_{2}^{+}$, H$^{-}$, D, D$^{+}$) coupled to a table of metal cooling rates computed under the assumption of collisional ionization equilibrium with the 2013 release of the \texttt{Cloudy} photo-ionization code \citep{2013RMxAA..49..137F}. The photo-heating, -ionization, and -dissociation rates from the radiation transport are coupled directly to the non-equilibrium primordial chemistry network. As mentioned earlier, we do not include dust grains and instead assume all metal remains in the gas phase, in contrast to \citet{2015MNRAS.452.2822S}.

Finally, we model the formation and feedback of individual Pop III stars using the method described in \citet{2012ApJ...745...50W}. We insert a star particle representing a 40 \msun\ Pop III star when the following criteria are satisfied: the proper baryon number density is greater than 10$^{7}$ cm$^{-3}$; the velocity field at the grid cell has negative divergence; the molecular hydrogen mass fraction (H$_{2}$ plus H$_{2}^{+}$) is greater than $5\times10^{-4}$ with respect to the total density; and the metallicity is less than 10$^{-4}$ \zsun. After a Pop III star particle is created, the radiation transport solver \citep{2011MNRAS.414.3458W} propagates H, He, and He$^{+}$ ionizing radiation, discretized into energy groups of 28 eV, 30 eV, and 58 eV, respectively. The H$_{2}$ photo-dissociating Lyman-Werner (LW) radiation is assumed to be optically thin. Its intensity is scaled as $r^{-2}$ from the source and decremented by H$_{2}$ self-shielding following \citet{2011MNRAS.418..838W}. The 40 \msun\ star particle has properties derived from the stellar models of \citet{2002A&A...382...28S}: a main-sequence lifetime of 3.86 Myr and photon luminosities of $2.47\times10^{49}$ s$^{-1}$ (28 eV), $1.32\times10^{49}$ s$^{-1}$ (30 eV), $8.80\times10^{46}$ s$^{-1}$ (58 eV), and $2.90\times10^{49}$ s$^{-1}$ (LW). After the star's lifetime, it explodes as a standard Type II core-collapse supernova with energy of 10$^{51}$ erg, metal yield of 11.19 \msun, and total ejecta mass of 38.6 \msun\ \citep{2006NuPhA.777..424N}. We emulate an explosion by depositing the energy and mass into a sphere of radius 10 proper pc. When gas meets the first three formation conditions but exceeds the threshold metallicity, we instead use the AMR to follow the gravitational collapse to densities at which our chemistry assumptions break down, then end the simulation.

\subsection{Simulation Setup}
\label{sec:setup}

This is a cosmological simulation with a 500 kpc/h comoving box size following the formation of a dark matter halo reaching $\sim$1.7$\times10^{7}$ \msun\ by a redshift of 10. The simulation is initialized at $z = 180$ with the \texttt{MUSIC} initial conditions generator \citep{2011MNRAS.415.2101H} with the \textit{WMAP 7} best-fitting cosmological parameters: $\Omega_{\rm m} = 0.266$, $\Omega_{\rm \lambda} = 0.732$, $\Omega_{\rm b} = 0.0449$, H$_{0} = 71.0$ km/s/Mpc, $\sigma_{8} = 0.801$, and n$_{\rm s} = 0.963$ \citep{2011ApJS..192...18K}, with an \citet{1999ApJ...511....5E} transfer function and using second-order perturbation theory. In an exploratory dark matter-only simulation with 512$^{3}$ particles, we identify our target halo as that with the most unique progenitor halos with masses of at least 10$^{6}$ \msun, thereby maximizing opportunities for Pop III star formation. We then regenerate the initial conditions with two levels of telescoping refinement surrounding the Lagrangian region of the target halo. This corresponds to an effective resolution of $2{,}048^{3}$ particles and a dark matter and baryon mass resolution of 1.274 \msun\ and 0.259 \msun, respectively.

As the simulation evolves we allow AMR within the high resolution region when any of the following criteria are met: there are more than four dark matter particles within a cell; the baryon mass exceeds four times the mean baryon mass per cell multiplied by a factor of 2$^{-0.2L}$, where $L$ is the refinement level (making baryon refinement slightly super-Lagrangian); or the local Jeans length is resolved by fewer than 64 cells. This Jeans length resolution has been shown to be sufficient to avoid numerical fragmentation \citep{2011ApJ...731...62F, 2014ApJ...783...75M}. The above criteria are given in roughly the order in which they become relevant. That is, refinement is first triggered by dark matter overdensity as a halo assembles, then baryon overdensity as gas settles into the potential well, and then by the Jeans length criterion when the gas becomes self-gravitating \citep{2014ApJ...783...75M}. We place no practical limit on the number of AMR levels and instead allow the simulation to proceed at the numerically appropriate resolution given the refinement criteria. In practice, the ``cruising altitude'' of the simulation is generally 8--9 levels of AMR ($\sim$3-5 pc comoving spatial resolution), reaching around 15 levels (0.04 pc comoving) just prior to Pop III star particle creation.

\section{The Evolution of Gas in Minihalos}
\label{sec:minihalos}

The model presented here is motivated by the behavior observed in the simulation of metal-free gas collapsing within the minihalos that go on to form Pop III stars. We first describe the evolution observed in the simulation, then present a simple model capable of reproducing the evolution of the central gas. Finally, we use this model to study a scenario in which minihalos are exposed to sufficient photo-dissociating radiation to totally destroy their molecular gas.

\subsection{From the Simulation}

Nine distinct Pop III star formation events occur within the simulation by the final output at $z \sim 11.8$. Of these, three are doubles (for a total of 12 stars) wherein two star particles are created in the same molecular cloud within 1000 years of each other. We do not examine further in this study the nature of these double events. Following the convention of \citet[][see Table 1]{2024MNRAS.527..307C}, we refer to each halo according to the star particle(s) created therein. Only seven of the nine halos are completely devoid of metals. The other two halos (3 and 12) are enriched by nearby supernovae. The metallicity of the gas in which their star particles form is below the threshold we use to consider them Pop II, hence they form as Pop III stars. Nevertheless, we omit these two halos from the study.

\begin{figure*}
    \includegraphics[width=\textwidth]{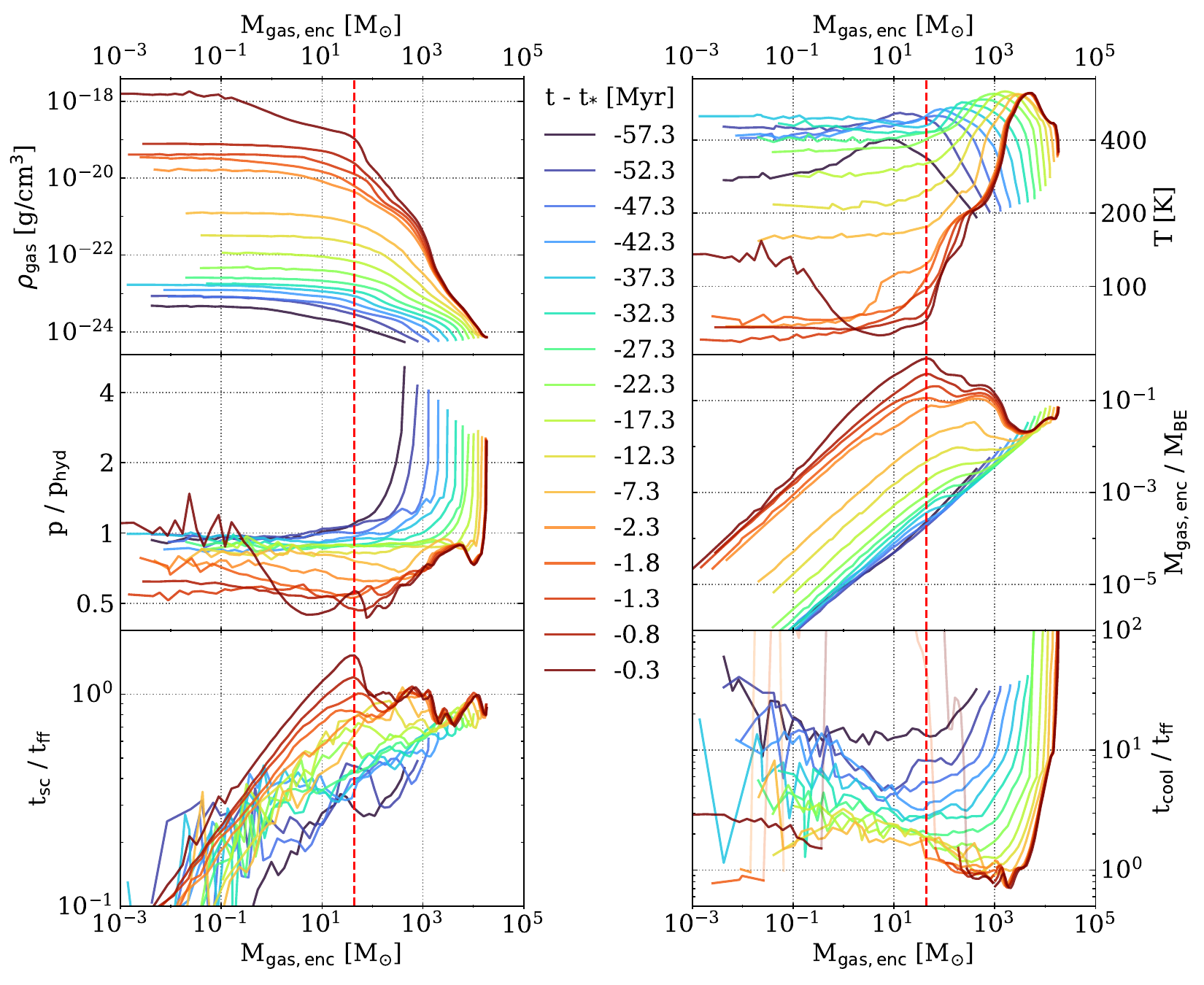}
    \caption{Radial profiles of first Pop III star-forming minihalo (halo 1) prior to the formation of the star particle. The outermost extent of the profile corresponds to the gas mass contained within the virial radius at that time. The panels show the evolution of gas density (upper-left), temperature (upper-right), ratio of gas pressure to hydrostatic equilibrium (middle-left), ratio of local Bonnor-Ebert mass to enclosed gas mass (middle-right), ratio of sound-crossing time to free-fall time (lower-left), and ratio of cooling time to free-fall time (lower-right, with opacity of the lines decreased to aid visibility where $t_{cool}$ becomes large when the cooling rate drops significantly). The center of each profile is calculated by starting with the baryonic center of mass for the entire halo, then iteratively decreasing the radius by 10\% and recomputing the center of mass until the sphere has reached 4$\times$ the minimum cell size. The profiles are then remapped to display quantities as a function of enclosed gas mass. Line colors correspond to the time before star particle creation, ranging from $\sim$60 Myr (dark blue) to $\sim$300 kyr (dark red, the last snapshot available). The vertical, dashed, red line denotes the most gravitationally unstable mass coordinate (i.e., the peak in Figure \ref{fig:BE-profiles}).}
    \label{fig:profiles}
\end{figure*}

We trace the merger history of each of the seven halos back to the point at which the most massive progenitor has a mass of $10^{4}$ \msun. Figure \ref{fig:profiles} shows radial profiles of the first halo to form a Pop III star. Being the first halo to undergo star formation, its gas is unaffected by radiative or supernova feedback from neighbors and it is thus the simplest case to display. We show the evolution of gas within the halo for the $\sim$60 Myr leading up to the creation of the star particle, which occurs at $z \sim 24$. The final profile shown corresponds to the last snapshot before the star particle was created. Over this time period the halo grows from a mass of $\sim$$1.1\times10^{4}$ \msun\ to $\sim$$1.2\times10^{5}$ \msun.

The central $\sim$10\% of the halo's gas is nearly in hydrostatic equilibrium for most of its evolution. Only in the final 7 Myr, when the temperature drops rapidly due to cooling by HD, is the central gas under-pressured with respect to hydrostatic equilibrium. The Bonnor-Ebert mass is the minimum mass in which a cored, isothermal sphere of gas experiencing a fixed external pressure becomes gravitationally unstable. The middle-right panel of Figure \ref{fig:profiles} shows the ratio of the enclosed gas mass to the local Bonnor-Ebert mass, given by
\begin{equation}
\label{eqn:mbe}
M_{{\rm BE}} \simeq 0.8 \times \frac{c_{s}^{4}}{G^{3/2} p^{1/2}},
 \end{equation}
where $c_{s}$ is the local sound speed, $G$ is the gravitational constant, and $p$ is the external pressure. Here, we approximate the external pressure as the maximum of the local thermal pressure and the hydrostatic pressure (i.e., the pressure exerted by the weight of the gas at larger radii, Equation \ref{eqn:phyd}). At an enclosed gas mass of roughly 40 \msun\ (denoted by the vertical, dashed line in Figure \ref{fig:profiles}) this ratio grows quickly at late times, approaching unity as the time of star particle creation nears. This formalism as we have employed it does not include the influence of the dark matter, but we can do so approximately by considering the ratio of the sound-crossing time and the free-fall time, which is expressed as
\begin{equation}
\label{eqn:tff}
t_{{\rm ff}} = \sqrt{\frac{3 \pi}{32 G \rho_{{\rm tot}}}},
\end{equation}
where $\rho_{{\rm tot}}$ is the sum of the gas and dark matter densities. This is the criterion for the Jeans mass, which applies to a parcel of gas at constant density and temperature. We show this in the lower-left panel of Figure \ref{fig:profiles}. At the 40 \msun\ mass coordinate, the sound-crossing time is lower than the free-fall time until late times. At early times the presence of the dark matter keeps the two timescales within a factor of a few, with the cooling time much longer than the free-fall time. This is consistent with the notion of quasi-hydrostatic evolution evidenced by the pressure profiles. Later, when the sound-crossing and free-fall times are similar, the gas appears to undergo runaway collapse. During this period, the cooling time becomes very long with respect to the free-fall time. This is due to the steep drop-off in the cooling rate as a function of temperature.


The peak in ratio of enclosed gas mass to Bonnor-Ebert mass occurs at the same mass coordinate as the peak in the ratio of the sound-crossing time to free-fall time. At this point the gas density dominates over that of the dark matter, so $\rho_{gas} \approx \rho_{tot}$. As well, if one takes the external pressure to be the local pressure, as we have done in Equation \ref{eqn:mbe}, then $c_{s}/\sqrt{p}$ reduces to $\rho^{-1/2}$, giving the Bonnor-Ebert and Jeans masses the same proportionality and a normalization constant differing only by a factor of $\sim$1.5.

\begin{figure}
    \includegraphics[width=\columnwidth]{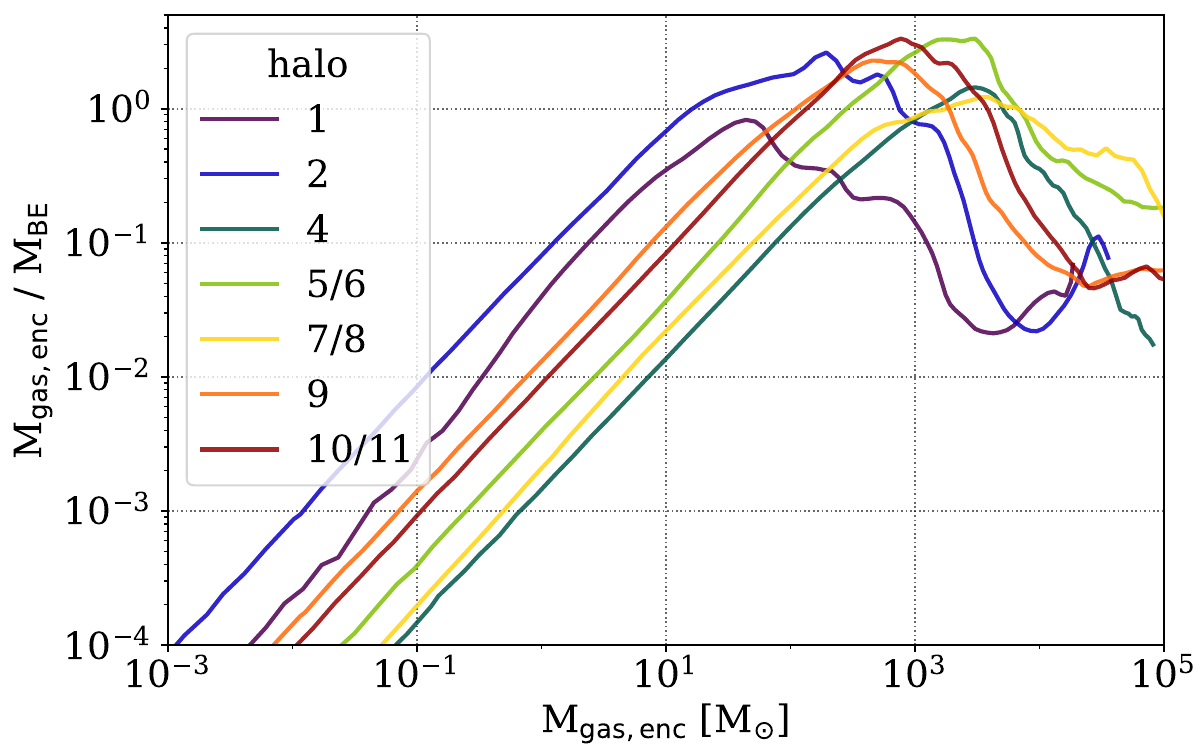}
    \caption{Profiles of the ratio of enclosed gas mass to local Bonnor-Ebert mass as a function of enclosed gas mass in the last snapshot prior to star particle creation for the seven halos modeled in this work. The third and ninth star-forming halos are omitted for not being entirely metal-free.}
    \label{fig:BE-profiles}
\end{figure}

We observe the behavior described above in all Pop III star-forming halos within the simulation. The central gas remains in approximate hydrostatic equilibrium as it slowly cools and condenses, eventually becoming under-pressured with respect to hydrostatic equilibrium at late times. The primary difference between each of the Pop III star-forming halos is the mass coordinate at which the Bonnor-Ebert mass is finally exceeded (Figure \ref{fig:BE-profiles}, i.e., the mass of gas that becomes gravitationally unstable). This ranges from about 40 \msun\ in halo 1 to nearly 1,600 \msun\ in halo 4, similar to the range observed in the sample of 100 minihalos simulated by \citet{2014ApJ...781...60H}. The ratio of gravitationally unstable cloud mass to halo mass shows a weakly positive trend with halo mass (albeit, in our very small sample), with a mean value of roughly 10$^{-3}$. The correlation between cloud mass and halo mass is not surprising as the Bonnor-Ebert mass increases with temperature, which in turn increases with virial mass (i.e., $T_{v} \propto M_{v}^{2/3} (1+z)$).

\subsection{The Minihalo Model}
\label{sec:model}

What drives the gas in these halos to condense, collapse, and eventually form stars? Our aim is to model the evolution of gas in each minihalo from the time it was first resolved in the simulation, taking into account their unique growth histories and external influences. In particular, we wish to follow the parcel of gas that ultimately becomes gravitationally unstable. Hence, it is crucial to reproduce the temporal evolution of the gas as well as its path through phase space (i.e., density vs. temperature). Below we present our one-zone Minihalo model, which takes into account the physical conditions experienced by individual halos and augments conventional free-fall evolution with a pressure-driven component. We first describe the model in full, then evaluate the importance of various physical processes.

\subsubsection{Model Summary}

As we wish to model the gas in each of our simulated minihalos individually, we must first construct the time-dependent data characterizing the influences experienced as the halos grow and other stars form nearby. We create radial profiles of each halo from every simulation snapshot in which it exists and remap them from radius to enclosed gas mass (e.g., Figure \ref{fig:profiles}). We identify the mass coordinate with the highest ratio of enclosed mass to Bonnor-Ebert mass in the snapshot immediately prior to Pop III star particle formation. We then initialize the relevant gas quantities (i.e., density, internal energy, chemical species fractions) using the values in that mass coordinate at the time when the halo's most massive progenitor has a virial mass of roughly 10$^{4}$ \msun. We also make note of the radius corresponding to the target mass coordinate in the profile. This serves as the initial radius of the gas parcel, which we update as the density changes, assuming spherical symmetry. As the model runs we update the dark matter density (used in the modified free-fall component) and photo-chemical/heating rates (stimuli to the chemistry solver) using the radial profiles at the relevant cosmic time and mass coordinate. From one time step to the next, we update the internal energy and chemical species using the \texttt{Grackle}\footnote{https://grackle.readthedocs.io/} chemistry and cooling library \citep{2017MNRAS.466.2217S} with the same settings as were used in the simulation.

Once the model is initialized, the procedure is as follows:
\begin{enumerate}
    \item Calculate time step as the minimum of the cooling time and free-fall time multiplied by a safety factor, here 0.01.
    \item \label{step:grackle} With current density ($\rho_{n}$), internal energy ($e_{n}$), and chemical species fractions ($\mathbf{y}_{n}$), solve chemistry and cooling to update $e_{n}$ and $\mathbf{y}_{n}$ to $e_{n+1}$ and $\mathbf{y}_{n+1}$ with $\rho$ unchanged.
    \item \label{step:rhoupdate} Update $\rho_{n}$ to $\rho_{n+1}$ following the process described in Section \ref{sec:density} below.
    \item Update the internal energy to account for the change in density from step \ref{step:rhoupdate} using the energy equation with the radiative cooling term removed (as it was applied in step \ref{step:grackle}). This is expressed as
    \begin{equation}
        \frac{de}{dt} = -p \frac{d}{dt} \frac{1}{\rho}.
    \end{equation}
    \item Update the cloud radius assuming conservation of mass (i.e., $\rho_{n} r_{n}^{3} = \rho_{0} r_{0}^{3}$).
\end{enumerate}
The model iterates these steps until the Bonnor-Ebert mass is exceeded, at which time we consider it to have entered runaway collapse. For the time-dependent inputs into the model (e.g., dark matter density, photo-ionization/heating rates, etc.), we perform a two-dimensional linear interpolation over time and radius from the radial profiles of the given halo.

\subsubsection{Density Evolution}
\label{sec:density}

At step \ref{step:rhoupdate}, we first determine the method of density update by comparing the sound-crossing time ($2r/c_{s}$) to free-fall time including the dark matter (i.e., Equation \ref{eqn:tff}). If the free-fall time is shorter we evolve the gas density according to the free-fall model of \citet{2005ApJ...626..627O}, which they modify to include the effect of pressure gradients acting to slow collapse (Equations 6--9 of that work). As we employ this in the model presented here, we outline the approach below. The evolution of the gas density in modified free-fall is expressed as
\begin{equation}
\label{eqn:model-density}
\frac{d \rho}{dt} = \frac{\rho}{t_{col}},
\end{equation}
where the collapse time, $t_{col}$, is altered from pure free-fall (i.e., Equation \ref{eqn:tff}) as
\begin{equation}
t_{col} = \frac{t_{ff}}{\sqrt{1-f}},
\end{equation}
and $f$ is the ratio of the force due to the local pressure gradient and gravity. This factor is approximated as a function of the effective adiabatic index ($\gamma \equiv \frac{\partial \log p}{\partial \log \rho}$) as
\begin{equation}
\label{eqn:ffactor}
f = \left\{
\begin{array}{lr}
0, & \gamma < 0.83,\\
0.6 + 2.5(\gamma - 1) - 6.0(\gamma - 1)^2, & 0.83 < \gamma < 1,\\
1.0 + 0.2(\gamma - 4/3) - 2.9(\gamma - 4/3)^2, & \gamma > 1.
\end{array}
 \right.
\end{equation}
As $\gamma \rightarrow 4/3$ and $f \rightarrow 1$, the gas is internally pressure supported and the collapse is halted. \citet{2005ApJ...626..627O} note that the evolution now depends on continued accretion of the surrounding material. They model this phase by setting an upper limit for $f$ of 0.95. In this work we allow $f$ to reach unity, however, at which time we consider the collapse to be fully stalled, thereby entering the pressure-driven mode. We evaluate the consequences of various choices for the value of $f$ below in Section \ref{sec:physics}.

When the sound-crossing time is shorter than the free-fall time or when $f = 1$ in the modified free-fall model, the density is updated according to a pressure-driven model. In this mode, the change in density between two time steps is given by
\begin{equation}
\label{eqn:pdrive}
\frac{\rho_{n+1}}{\rho_{n}} = \frac{p_{n+1}}{p_{n}} \frac{\mu_{n+1}}{\mu_{n}} \frac{T_{n}}{T_{n+1}}.
\end{equation}
The variables, $p$, $\mu$, and $\rho$, with ``$n$'' subscripts represent the thermal pressure, mean molecular weight, and gas density from the previous time step. Given a time step, $dt$, we integrate the chemistry and cooling starting with the gas described by the $n$-subscript variables. This integration evolves the chemical species and internal energy at a constant density, $\rho_{n}$. After this time step, the temperature, $T_{n+1}$, and mean molecular, $\mu_{n+1}$, are calculated from the updated species densities and internal energy. The new density, $\rho_{n+1}$, is then calculated following Equation \ref{eqn:pdrive}. The key variable driving the system is, thus, $p_{n+1}$, which we take to be the maximum of the thermal pressure and the hydrostatic pressure at the new time and radius as derived from the radial profiles of the halo. The hydrostatic pressure is given by
\begin{equation}
\label{eqn:phyd}
P_{hyd}(r) = \int_{R}^{r} \frac{G m_{tot, enc}}{r^{2}} \rho_{gas}\ dr,
\end{equation}
where $R$ is the outer radius (here taken to be the virial radius) and $m_{tot, enc}$ is the total (gas and dark matter) enclosed mass. Here, we label the gas density as $\rho_{gas}$ to emphasize that while the dark matter and gas act in tandem to pull material down toward the center, only the gas weighs down on the material from above (i.e., the outside).

\subsubsection{The Importance of Gravity, Pressure, and Radiation}
\label{sec:physics}

\begin{figure}
    \includegraphics[width=\columnwidth]{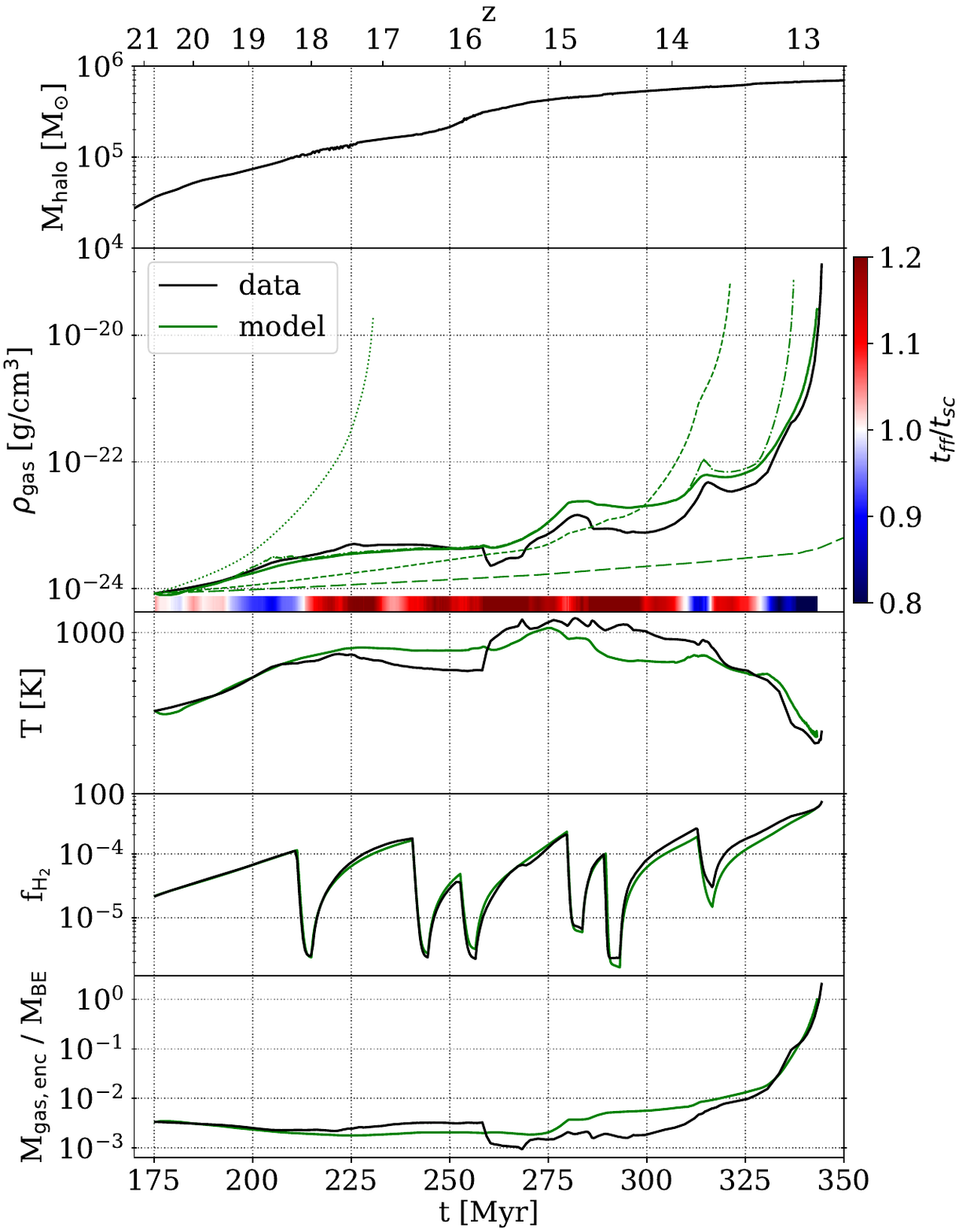}
    \caption{Comparison of the one-zone Minihalo model with the simulation. The evolution of halo mass, gas density, temperature, H$_{2}$ fraction, and Bonnor-Ebert mass ratio for halo 10/11. The curves represent the mass coordinate with the highest ratio of enclosed mass to Bonnor-Ebert mass immediately prior to star formation (i.e., Figure \ref{fig:BE-profiles}). Black lines show the values extracted from the simulation and solid, green lines are the full one-zone model. The colored band indicates the ratio of the free-fall to sound-crossing time and hence the mode of the model, with red denoting the pressure-dominated and blue the modified free-fall. The four broken lines in the density panel illustrate variants of a free-fall model. Three ignore the pressure-driven component: \textit{dotted}, pure free-fall; \textit{dashed}, modified free-fall with $f$ capped at 0.95; \textit{long dashed}, same with $f$ capped at 0.99. Finally, \textit{dash-dot}, hydrostatic pressure with pure free-fall.}
    \label{fig:control-model-8}
\end{figure}

In Figure \ref{fig:control-model-8}, we show the results of the full model, alongside several variants, for halo 10/11. This halo, the last in our sample to form a star, has been influenced by the seven prior instances of star formation (the earliest of which does not appear in the figure) and, thus, represents one of the more complex cases. The Minihalo model reproduces the time to runaway collapses to within about 1 Myr over a total time of 169 Myr. Gravity alone would drive the system to collapse on the timescale proportional to Equation \ref{eqn:tff}. In the case of halo 10/11, the initial gas and dark matter densities are $8\times10^{-25}$ g/cm$^{3}$ and $8\times10^{-24}$ g/cm$^{3}$, respectively, corresponding to a free-fall time of about 22 Myr and an overall collapse time about twice that (i.e., the dotted, green line in Figure \ref{fig:control-model-8}).

What is also striking is the degree to which just six brief episodes of star formation (3.86 Myr each) delay the ultimate onset of free-fall by more than 100 Myr. At $z \sim 19$, the gas density is evolving via the free-fall mode until a Pop III star forms just prior to $z \sim 18$. At this point the halo enters a cycle of seeing its molecular content recover and be destroyed again \citep{2024MNRAS.527..307C}, all the while evolving in the pressure-driven mode. The white bands that appear in Figure \ref{fig:control-model-8} during this period illustrate how close the halo comes to breaking out of this cycle on multiple occasions as f$_{\rm H_{2}}$ exceeds 10$^{-4}$. Another key to this is the relatively low gas density, which prevents effective self-shielding. Further inside the halo where densities are higher, self-shielding is certainly taking place, but this is ultimately unimportant as it is this mass coordinate which determines when star formation occurs in this halo.

Thinking back to the far simpler case of halo 1, where Figure \ref{fig:profiles} shows evolution over a 57 Myr period, the initial proper total density at the relevant mass coordinate (here, $\sim$40 \msun) is roughly $2\times10^{-22}$ g/cm$^{3}$, corresponding to a free-fall time of just under 5 Myr. However, it is apparent from the pressure profiles that the gas is evolving very close to hydrostatic equilibrium until just about 7 Myr before final collapse. This phenomenon is also observed at the extreme opposite end of the halo mass spectrum, where hydrostatic equilibrium accounts for up to 95\% of the gas pressure in galaxy clusters \citep{2009ApJ...705.1129L}. What links the scenarios is the ineffectual radiative cooling, which here is due to H$_{2}$ and in galaxy clusters is offset by feedback from active galactic nuclei \citep{2015Natur.519..203V}.

Regardless, it is clear that thermal pressure is important for regulating the rate of collapse. In their one-zone models, \citet{2005ApJ...626..627O} account for this by modifying the free-fall collapse model introduced in \citet{2000ApJ...534..809O} to slow the density evolution by a factor relating to the ratio of pressure gradient forces to gravity (i.e., Equations \ref{eqn:tff}, \ref{eqn:model-density}--\ref{eqn:ffactor}), and setting a maximum value of $f = 0.95$. For the halos in this work this generally results in collapse proceeding too quickly, as illustrated by the dashed line in Figure \ref{fig:control-model-8}. Increasing the upper limit on $f$ to 0.99 further slows collapse, but in the case of halo 10/11 (Figure \ref{fig:control-model-8}, long dashed line) by too much, whereas in other cases it is still not enough. It is also clear from Figure \ref{fig:control-model-8} that this approximation fails to reproduce the low density, pressure-dominated evolution entirely\footnote{It should be noted that, while upper limits on $f$ greater than 0.95 resulted in collapse times both too short and too long, a value of 0.95 was fairly consistently too short and likely the best value to use.}.

Finally, despite the improvement made by the addition of hydrostatic pressure, we note that slowing free-fall by pressure-gradient forces is still a necessary component for the late stages of the evolution, as is illustrated by the dash-dot line in Figure \ref{fig:control-model-8}. This term could be implemented more directly as essentially the derivative of Equation \ref{eqn:phyd}, but for simplicity we choose to stay with the approximation of Equation \ref{eqn:ffactor} for this work.

\subsection{A Critical Metallicity for Star Formation}
\label{sec:model-result}

We have constructed a model that describes reasonably well the behavior of gas in the interiors of minihalos. We now undertake a thought experiment and ask how these halos would evolve in the total absence of H$_{2}$ resulting from a strong UV background or repeated nearby star formation. In the metal-free limit the gas would not be able to cool until its halo's virial temperature reaches $\sim$10$^{4}$ K at a mass of roughly 10$^{7}$ \msun. How much metal is required to supplement the loss of H$_{2}$ and induce gravitational instability before the halo reaches the atomic cooling limit?

We try to answer this question in the simplest way possible, by deactivating the H$_{2}$-related chemistry altogether. This requires the Minihalo model to run past the point in which each of the halos originally formed its Pop III star. After this point in the simulation, the halo is evacuated of its baryon content by the stellar radiation and subsequent supernova. To approximate how the hydrostatic pressure would have evolved if no star had formed, we assume the gas density profile of the interior of the halo remains fixed in the state just prior to star formation. We continue to sample the dark matter profile of the halo from the simulation. As the halo grows in dark matter mass we assume that it accretes a mass of baryons proportional to the cosmic ratio. For the purposes of the model we deposit the accreted baryons into the outermost bin (i.e., the current virial radius) of the gas density profile to act as additional weight on the inner gas. As significant as these assumptions are, we find that the model behaves fairly sensibly in the limit of no metals or molecules (H$_{2}$/HD). The density and temperature both increase steadily as the halo grows, and the ratio of enclosed mass to Bonnor-Ebert mass decreases to $\sim$10$^{-4}$. For each halo we measure the time required for the gas temperature in the metal-free control run to reach 8000 K, where cooling from H Lyman-$\alpha$ starts to become significant. We set this as the maximum run time for the models discussed next. Beyond this, the minihalos will grow into atomic cooling halos where H$_{2}$ and star formation are significantly less inhibited by external UV radiation \citep{2002ApJ...569..558O}.

To this point, we have focused on the evolution of the mass coordinate of halo gas which becomes the most gravitationally unstable. However, the characteristic mass of gravitational instability is likely to change as we increase the metallicity \citep{2022MNRAS.509.1959S}. To account for this, we modify the Minihalo model to simultaneously follow a contiguous range of mass coordinates, ranging from 5 \msun\ to twice the original target coordinate of each halo (i.e., Figure \ref{fig:BE-profiles}). For our purposes this only involves modifying the hydrostatic pressure calculation to account for the change in position of the modeled mass coordinates. We confirm that this variant of the model reproduces the behavior of each of the mass coordinates modeled individually using the original one-zone version.

\begin{figure}
    \includegraphics[width=\columnwidth]{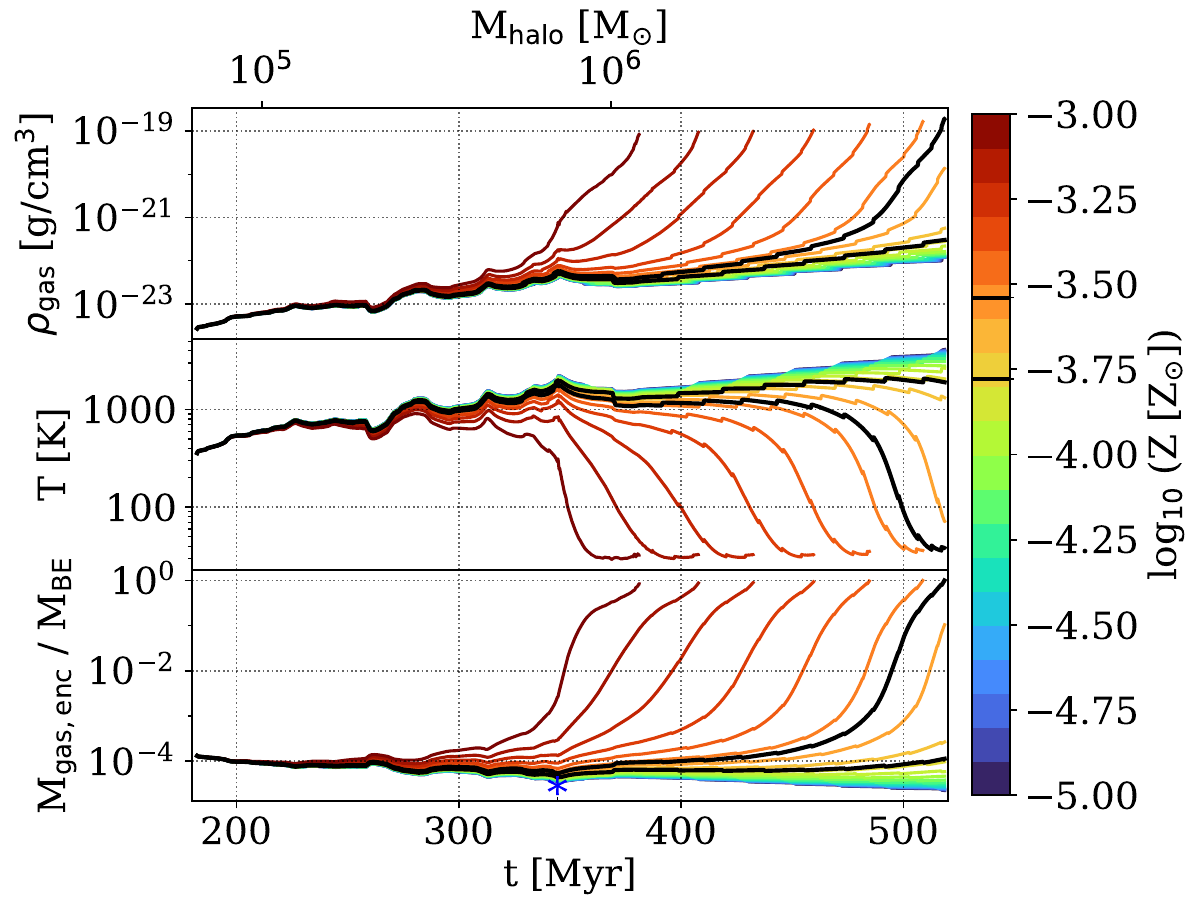}
    \caption{Density (top), temperature (middle), and Bonnor-Ebert ratio (bottom) from the expanded Minihalo model for halo 10/11 with varying metallicity and primordial molecular chemistry disabled. The black lines denotes the critical and minimum metallicities for this halo. The blue asterisk on the bottom denotes when Pop III star forms in the original simulation.}
    \label{fig:metallicity-model-8}
\end{figure}

At last, we run the expanded model with the primordial molecular chemistry disabled and gradually increase the metallicity. We use the tabulated metal cooling described in \citet{2008MNRAS.385.1443S}, which assumes a Solar abundance pattern and collisional ionization equilibrium. The results of this experiment are shown for halo 10/11 in Figure \ref{fig:metallicity-model-8}. At a metallicity of 10$^{-5}$ \zsun the gas evolves similarly to the metal-free case. Prior to the onset of Lyman-$\alpha$ cooling, the gas is actually more stable against collapse than at the start of the model. At higher metallicities, the gas is able to cool and become gravitationally unstable. For each halo, we calculate a ``critical metallicity,'' $\mathrm{Z_{cr}}$, as the minimum metallicity required for at least one mass coordinate to exceed the Bonnor-Ebert mass by the maximum run time. We also calculate a ``minimum metallicity,'' $\mathrm{Z_{min}}$, as that required for the time derivative of the Bonnor-Ebert ratio for at least one mass coordinate to become positive. This is the metallicity at which cooling is just beginning to counterbalance the compression from hydrostatic pressure. These two metallicity values are illustrated as the black lines in Figure \ref{fig:metallicity-model-8}. We present the values for the critical and minimum metallicities for all halos in Table \ref{tab:Zcr}. We find mean values of [Z$_{\rm cr}$] = -3.71 $\pm$ 0.16 and [Z$_{\rm min}$] = -4.04 $\pm$ 0.23.

Finally, we observe a sharp transition in the most gravitationally unstable mass coordinate. At low metallicities it is always the largest mass coordinate with the highest Bonnor-Ebert ratio at the end of the model, albeit significantly less than unity. In each halo this drops abruptly to roughly 10s of \msun\ between the minimum and critical metallicities, then increases gradually with metallicity as more of the gas is able to reach the CMB temperature. This is similar to the behavior reported by \citet{2022MNRAS.509.1959S}, although with significant differences in the physical setup and details of the results. We mention this result mainly for completeness, but caution against further interpretation as the nature of the stellar IMF is not the focus of this work.

\begin{table}
\centering
\caption{The critical, [Z$_{\rm cr}$], and minimum, [Z$_{\rm min}$], metallicities for each halo in the Minihalo model, with halos numbered according to \citet{2024MNRAS.527..307C}. The metallicities shown are log$_{10}$ with respect to the Solar metallicity, where log$_{10}$(\zsun) = -1.89. The abundance pattern of the cooling table has [C/H] = -3.61 and [O/H] = -3.31, corresponding to $D_{\rm trans}$ = [Z] + 0.11 \citep{2007MNRAS.380L..40F}. The final two rows give the mean and standard deviation for the sample. }
\label{tab:Zcr}
\begin{tabular}{lrr} 
\hline
Halo & [Z$_{\rm cr}$] & [Z$_{\rm min}$]\\
\hline
1 & -3.91 & -4.21 \\
2 & -3.89 & -4.10 \\
4 & -3.66 & -4.25 \\
5/6 & -3.54 & -3.94 \\
7/8 & -3.59 & -3.67 \\
9 & -3.86 & -4.35 \\
10/11 & -3.54 & -3.78 \\
\hline
mean & -3.71 & -4.04 \\
std. & 0.16 & 0.23 \\
\hline
\end{tabular}
\end{table}

\section{The Metallicity Floor and CEMP Stars}
\label{sec:interpretation}

\begin{figure}
    \includegraphics[width=\columnwidth]{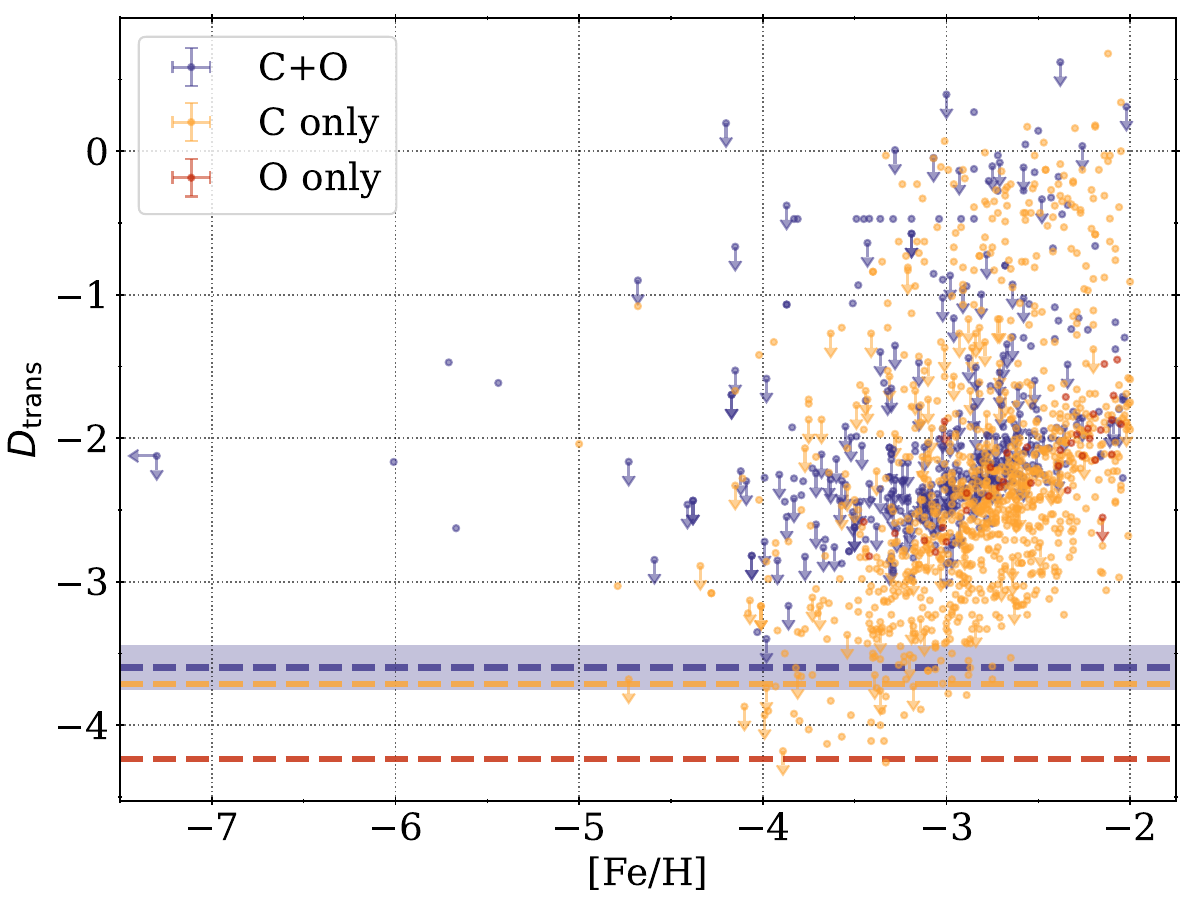}
    \caption{The transition discriminant, $D_{\rm trans}$, as a function of [Fe/H] for all metal-poor stars with [Fe/H] $<$ -2 in the JINAbase metal-poor star database \citep{2018ApJS..238...36A}. The blue points denote stars where both C and O abundances have been measured while yellow and red points correspond to stars which only C or O, respectively, have been measured. The blue dashed line and shaded region correspond to our estimate of $D_{\rm trans,crit}$ and one sigma uncertainty. The yellow and red dashed lines show $D_{\rm trans,crit}$ for when abundances are known for only C or O, respectively, assuming a Solar abundance pattern, i.e., the given star has the same abundances relative to Solar for both C and O.}
    \label{fig:Dtrans}
\end{figure}

The goal of this work has been to understand the conditions which lead to the formation of star-forming gas in minihalos. We have refined one-zone models based on free-fall collapse to match the evolution observed in cosmological simulations. In doing so, we have established the importance of hydrostatic pressure in driving the gradual increase in gas density as the halo grows. We have also confirmed the validity of the modifications made by \citet{2005ApJ...626..627O} to the free-fall model to approximate the slowing of collapse by pressure gradient forces.

The real question we are seeking to answer, however, is how would these minihalos behave if they grew up in the early formation environment of the Milky Way? Such a large-scale overdensity would have a highly clustered population of Pop III stars and thus be replete with photo-dissociating radiation and metals, two powerful terms primarily acting in opposition. In the extreme limit of full suppression of H$_{2}$, what abundance of metal is required such that gravitationally unstable gas can still form? Taking the mean of the sample of minihalos studied here, we find this value to be roughly $10^{-3.7}$ \zsun\ for a Solar abundance pattern. This value is remarkably close to the apparent floor in C/O abundances observed in metal-poor stars, first quantified by \citet{2007MNRAS.380L..40F} as $D_{\rm trans,crit} = -3.5 \pm 0.2$.
Equation \ref{eqn:dtrans} derives from the fact that C {\sc ii} and O {\sc i} fine-structure emission dominate the metal cooling at low density, with the factor of 0.3 acting to normalize the cooling rates of the two terms. With the abundance pattern used here, our prediction of [Z$_{\rm cr}] = -3.71$ is equivalent to $D_{\rm trans,crit} = -3.60$. In Figure \ref{fig:Dtrans}, we show $D_{\rm trans}$ vs. [Fe/H] for all objects from the JINAbase metal-poor star database \citep{2018ApJS..238...36A} with [Fe/H] $<$ -2, with our estimate of $D_{\rm trans,crit}$ overplotted. Of the $1{,}463$ stars plotted, only 23 ($\sim$1.6\%) have $D_{\rm trans}$ values below our one-sigma lower bound prediction of -3.76. All of those 23 stars only have measured C abundances, making their $D_{\rm trans}$ values lower limits. For example, if these stars had [C/O] = 0 (i.e., a Solar ratio of C to O), their associated $D_{\rm trans}$ values would increase by 0.11. For reference, the mean value of [C/O] for the stars shown in Figure \ref{fig:Dtrans} with both C and O abundances is -0.46 $\pm$ 0.98. If the stars with unmeasured O had [C/O] = -0.46, this would increase their $D_{trans}$ values by 0.27.

The interpretation here is that the apparent $D_{\rm trans}$ floor traces the metallicity at which minihalos are no longer prevented from forming stars by Lyman-Werner radiation of any intensity. In a cosmological context, these objects would be the most metal-poor ``metal-cooling halos,'' halos just below the atomic cooling limit whose radiative cooling is dominated by gas-phase metals \citep{2014MNRAS.442.2560W}. This is a variation on the original notion of $D_{\rm trans}$ as the minimum metallicity at which fragmentation, and hence low mass star formation, can first occur. The similarity between our prediction and the fragmentation-based one \citep[e.g.,][]{2003Natur.425..812B, 2007MNRAS.380L..40F} derives from their common foundation of 
metal coolants supplementing H$_{2}$ at low densities (1 cm$^{-3} \lesssim n \lesssim 10^{4}$ cm$^{-3}$) and temperatures \citep[100 K $\lesssim T \lesssim$ 1000 K, ][]{2008MNRAS.385.1443S}. The primary obstacle for the fragmentation argument is the ability of dust to induce fragmentation at much lower metallicities \citep[e.g.,][]{2005ApJ...626..627O, 2012MNRAS.419.1566S}. Although we do not include dust in our model, this does not present an issue for our interpretation. The effects of dust (i.e., heat exchange with the gas and catalyst for H$_{2}$ formation) are negligible at low densities in low metallicity environments \citep{2000ApJ...534..809O}. As well, the minihalo interpretation has the added benefit of a degree of natural variability from halo to halo, as illustrated by Table \ref{tab:Zcr}, and also from variations in the local radiation field. Thus, it is fairly robust to a small number of exceptions. It should also be noted here that prompt external enrichment \citep{2015MNRAS.452.2822S} provides a mechanism for forming stars well below $D_{\rm trans,crit}$.

Our prediction bears some similarity to that of \citet{2019ApJ...870L...3H}, who argue that carbon-enhanced gas clouds would collapse more quickly than a relatively carbon-deficient neighbor, then form massive stars (along with low-mass stars) and suppress star formation in the neighbor. Following this, they predict that star formation would be prevented below a critical carbon abundance, [C/H] $< -3.6$. This, too, derives mainly from being roughly the crossover point between H$_{2}$ and metal line cooling. In both scenarios, the setting is an enriched minihalo bathed in H$_{2}$-dissociating radiation. The primary difference is the source of the radiation. In their model it is the newly-formed carbon-enhanced star, with some dependence on the stellar mass, distance, and density of the secondary cloud. In ours it is the radiation from distant Pop III stars. The two lines of reasoning are not mutually exclusive, although their scenario requires a particular configuration and also would not prevent lower metallicity carbon-normal (i.e., [C/Fe] = 0) stars from forming in the case where metals were mixed evenly. On the other hand, as we have shown in Section \ref{sec:physics}, even a handful of nearby, short-lived star formation events producing $J_{21}$ values of roughly 0.1 \citep{2024MNRAS.527..307C} can render H$_{2}$ cooling insufficient in small minihalos ($\lesssim10^{6}$ \msun) for long periods of time \citep[see also][]{2019MNRAS.488.3268A}. Within the overdense environment that would characterize the early history of the Milky Way, such a value should also be reasonable to achieve as a background \citep[e.g.,][]{2012MNRAS.419..718H, 2013MNRAS.428.1857J, 2013ApJ...773..108C, 2023MNRAS.522..330I, 2024ApJ...962...62F}. During this time, metals created by the stars that did form will steadily enrich these irradiated minihalos \citep{2011MNRAS.414.1145M, 2021ApJ...909...70H} to the required metallicity.

Another commonality with \citet{2019ApJ...870L...3H}, and any other cooling-based explanations for CEMP stars or a C/O abundance floor, is the need to suppress iron enrichment at low metallicity. \citet{2019ApJ...870L...3H} highlight two avenues for this: SN that inherently produce enhanced light element yields \citep[e.g.,][]{2014ApJ...791..116C, 2014ApJ...792L..32I, 2023MNRAS.526.4467J} and varied mixing of different elements created in more conventional explosions \citep[e.g.,][]{2016MNRAS.456.1410S}.
Regardless of the mechanism that drives the carbon enhancement, there is compelling evidence from chemical enrichment that minihalos play a unique role in forming the lowest metallicity stars. In particular, minihalos appear to be necessary for capturing the extreme iron-poor tail of the Milky Way MDF \citep{2017MNRAS.465..926D} and uniquely capable of hosting true second-generation and CEMP stars \citep{2015MNRAS.454..659J}. This second point originally relied on faint (low-energy) Pop III SNe where timely fallback onto minihalos was possible \citep{2014ApJ...791..116C}, but the discovery of prompt external enrichment \citep{2015MNRAS.452.2822S} has provided avenue for second-generation star formation from virtually any progenitor.

\section{Caveats}
\label{sec:caveats}

It is important to note that the experiment presented here is but a crude simulacrum of star formation in metal-enriched minihalos during the early history of the Milky Way. The precise critical metallicity (or $D_{\rm trans,crit}$) values we calculate are the product of several simplifications, which we outline here. Indeed, visual inspection of Figure \ref{fig:Dtrans} suggests our value could be slightly high for the observational data. Positive contributions (i.e., leading to cooler gas) that we have omitted or oversimplified will lower the required metallicity. Likewise, any negative contributions (i.e., sources of heat) that we may have underestimated will raise it. First and foremost, we have disabled H$_{2}$ formation entirely in the chemistry network as a means of approximating complete dissociation by a Lyman-Werner background. More realistically, H$_{2}$ will start to be effectively self-shielded from $J_{21}$ values of $\sim$1 for halos above 10$^{6}$ \msun\ \citep{2021ApJ...917...40K, 2021MNRAS.507.1775S}. We have similarly ignored the effects of dust grains, whose positive contributions include providing an additional channel for H$_{2}$ formation, a heat sink for the gas \citep{1979ApJS...41..555H}, and shielding H$_{2}$ from Lyman-Werner radiation \citep{2005MNRAS.356.1529H}. As pointed out recently by \citet{2022MNRAS.509.1959S}, however, the role of dust as a coolant for the gas (i.e., transferring heat by inelastic collisions and re-radiating it as thermal emission) is an exception confined to when the gas is hotter than the dust. Stellar radiation couples strongly to dust and can easily turn dust into a source of heat. While only FUV photons will lead to photoelectric heating from grains, dust can be heated by radiation at much lower energies \citep{2014MNRAS.437.1662K}. Lastly, the metal cooling we employ assumes collisional ionization equilibrium (i.e., no ambient radiation background), which is clearly at odds with the physical scenario. Incident radiation adds a heating term, but also alters the ionization balance in a crucial way. Specifically, far more of the cooling from carbon then comes from fine-structure lines of C {\sc ii} rather than C {\sc i}, as the C {\sc i} ionization energy is less than 1 Ryd. At low densities ($n \lesssim 10^{3}$ cm$^{-3}$) and temperatures in the hundreds of K, C {\sc ii} cooling is $\sim$10--50 times more efficient than C {\sc i} \citep{1989ApJ...342..306H}. However, at higher temperatures, where we first see metals beginning to counteract compression heating (i.e., Figure \ref{fig:metallicity-model-8}), cooling from O {\sc i} (whose ionization energy is greater than 1 Ryd) is far more important. A more sophisticated treatment combining non-equilibrium metal and dust chemistry with a radiation background appropriate to the Milky Way's early formation environment \citep[e.g.,][]{2015MNRAS.446..160A} is beyond the scope of this work. However, such an endeavor could establish the sensitivity of this critical metallicity to the intensity of the radiation field, and thus open another avenue for observations of metal-poor stars to tell us something about the origins of the Galaxy.

\section{Summary and Conclusions}
\label{sec:conclusion}

We present in this work a case that the metallicity floor observed in Galactic metal-poor stars is directly tied to the conditions that allow for star formation in the minihalos present in the early formation environment of the Milky Way. When combined with a mechanism for suppressing iron at low metallicity, this provides a natural explanation for the origin CEMP stars. We summarize the key points of this argument below:

\begin{enumerate}
  \item We analyze a cosmological simulation that follows an environment of co-evolving Pop III star-forming minihalos with a dark matter mass resolution of $\sim$1 \msun\ and radiation hydrodynamics. We characterize the evolution of the baryon content of these halos from their earliest history through the development of gravitational instability. Until very late times the central gas evolves in approximate hydrostatic equilibrium as it cools slowly through H$_{2}$ and HD. This period is easily prolonged by brief episodes of Pop III star formation that photo-dissociate the molecules within the mass coordinate that will eventually become gravitationally unstable. Peaks eventually emerge in radial profiles of the ratio of enclosed baryon mass to Bonnor-Ebert mass and generally align with peaks in the ratio of the sound-crossing time to free-fall time, albeit with different absolute values. The gas becomes under-pressured with respect to hydrostatic equilibrium when the sound-crossing time is roughly the free-fall time.
  
  \item Motivated by the simulation data, we construct a one-zone model for the evolution of gas in minihalo cores. A parcel of gas evolves according to pressure confinement when the sound-crossing time is shorter than the free-fall time, where the pressure is the maximum of the local thermal pressure and the hydrostatic pressure. When the opposite is true, the density follows a free-fall model modified to include an opposing force from thermal pressure gradients. We couple this model to radial profiles of the simulated minihalos to provide time-dependent hydrostatic pressure values and radiative input from nearby Pop III stars. The model is able to accurately reproduce the density and temperature evolution of core gas from an arbitrarily early time in the halo's history up to when the gas becomes Bonnor-Ebert unstable.

  \item We use the model to investigate how the simulated minihalos would evolve in the presence of a strong LW background by re-running them with molecular chemistry disabled. As expected, the gas that formed stars under normal conditions now remains stable as the halo grows to the atomic cooling limit. If we then add metals to supplement the loss of primordial coolants, we find on average that minihalos are able to form stars before reaching the atomic cooling limit when enriched to a critical metallicity, Z$_{\rm cr} \sim 10^{-3.71}$ \zsun.
\end{enumerate}

The original aim of the simulation presented here was to ascertain what, if any, relevance the gas-phase critical metallicity of ($\sim$10$^{-3.5}$ \zsun) has to the transition from Pop III to Pop II star formation. It has been fairly clear for some time that it is not related to fragmentation. That role belongs to the dust and its associated critical metallicity of $\sim$10$^{-5.5}$ \zsun. Is is vexing, though, why the dust-phase critical metallicity is not more apparent in observational data. For now, however, the story of Milky Way's oldest stars will remain a tale of two metallicities.
 
\section*{Software}

The simulation was initialized with the \texttt{MUSIC} initial conditions generator \citep{2011MNRAS.415.2101H}, run using the \texttt{Enzo} code \citep{2014ApJS..211...19B, 2019JOSS....4.1636B} and the \texttt{Grackle} chemistry and cooling library \citep{2017MNRAS.466.2217S}, and analyzed with \texttt{consistent-trees} \citep{2013ApJ...763...18B}, \texttt{ROCKSTAR} \citep{2013ApJ...762..109B}, \texttt{yt} \citep{2011ApJS..192....9T}, \texttt{yt\_astro\_analysis} \citep{yt-astro}, and \texttt{ytree} \citep{ytree, ytree-zenodo}. The main software dependencies were \texttt{Python 3} \citep{van1995python}, \texttt{NumPy} \citep{harris2020array}, \texttt{matplotlib} \citep{Hunter2007}, and \texttt{mpi4py} \citep{mpi4py1, mpi4py2, mpi4py3, mpi4py4, mpi4py5}.

\section*{Acknowledgements}

We thank the referee, Gerry Gilmore, for helpful comments on the manuscript. BDS wishes to thank Heather Jacobson, John Regan, and the entire FOGGIE collaboration (in particular, Molly Peeples) for useful conversations and continued support. BDS and SK are supported by STFC Consolidated Grant RA5496. BWO acknowledges support from NSF grants \#1908109 and \#2106575 and NASA ATP grants NNX15AP39G and 80NSSC18K1105. JHW acknowledges support from NSF grant \#2108020 and NASA grant 80NSSC20K0520. BWO and JHW acknowledge support from NASA TCAN grant 80NSSC21K1053. This research is part of the Blue Waters sustained-Petascale Computing Project, which is supported by the NSF (award number ACI-1238993) and the state of Illinois. Blue Waters is a joint effort of the University of Illinois at Urbana-Champaign and the National Center for Supercomputing Applications. The simulations described by this paper were run using an NSF PRAC allocation (award number OCI-0832662). For the purpose of open access, the author has applied a Creative Commons Attribution (CC BY) licence to any Author Accepted Manuscript version arising from this submission.

\section*{Data Availability}

All codes to produce the analysis and figures for this work are available in the \texttt{yt\_p2p}\footnote{https://github.com/brittonsmith/yt\_p2p} \texttt{yt} extension and the ``p2p'' branch of the lead author's fork of \texttt{Grackle}\footnote{https://github.com/brittonsmith/grackle/tree/p2p}. Data products used in this work include simulation snapshots, halo catalogs, merger trees, and halo radial profiles. These will be made available upon reasonable request to the lead author.



\bibliographystyle{mnras}
\bibliography{p2p2} 

\begin{thebibliography}{}
\makeatletter
\relax
\def\mn@urlcharsother{\let\do\@makeother \do\$\do\&\do\#\do\^\do\_\do\%\do\~}
\def\mn@doi{\begingroup\mn@urlcharsother \@ifnextchar [ {\mn@doi@}
  {\mn@doi@[]}}
\def\mn@doi@[#1]#2{\def\@tempa{#1}\ifx\@tempa\@empty \href
  {http://dx.doi.org/#2} {doi:#2}\else \href {http://dx.doi.org/#2} {#1}\fi
  \endgroup}
\def\mn@eprint#1#2{\mn@eprint@#1:#2::\@nil}
\def\mn@eprint@arXiv#1{\href {http://arxiv.org/abs/#1} {{\tt arXiv:#1}}}
\def\mn@eprint@dblp#1{\href {http://dblp.uni-trier.de/rec/bibtex/#1.xml}
  {dblp:#1}}
\def\mn@eprint@#1:#2:#3:#4\@nil{\def\@tempa {#1}\def\@tempb {#2}\def\@tempc
  {#3}\ifx \@tempc \@empty \let \@tempc \@tempb \let \@tempb \@tempa \fi \ifx
  \@tempb \@empty \def\@tempb {arXiv}\fi \@ifundefined
  {mn@eprint@\@tempb}{\@tempb:\@tempc}{\expandafter \expandafter \csname
  mn@eprint@\@tempb\endcsname \expandafter{\@tempc}}}

\bibitem[\protect\citeauthoryear{{Abohalima} \& {Frebel}}{{Abohalima} \&
  {Frebel}}{2018}]{2018ApJS..238...36A}
{Abohalima} A.,  {Frebel} A.,  2018, \mn@doi [\apjs]
  {10.3847/1538-4365/aadfe9}, \href
  {https://ui.adsabs.harvard.edu/abs/2018ApJS..238...36A} {238, 36}

\bibitem[\protect\citeauthoryear{{Agarwal} \& {Khochfar}}{{Agarwal} \&
  {Khochfar}}{2015}]{2015MNRAS.446..160A}
{Agarwal} B.,  {Khochfar} S.,  2015, \mn@doi [\mnras] {10.1093/mnras/stu1973},
  \href {https://ui.adsabs.harvard.edu/abs/2015MNRAS.446..160A} {446, 160}

\bibitem[\protect\citeauthoryear{{Agarwal}, {Cullen}, {Khochfar}, {Ceverino}
  \& {Klessen}}{{Agarwal} et~al.}{2019}]{2019MNRAS.488.3268A}
{Agarwal} B.,  {Cullen} F.,  {Khochfar} S.,  {Ceverino} D.,   {Klessen} R.~S.,
  2019, \mn@doi [\mnras] {10.1093/mnras/stz1347}, \href
  {https://ui.adsabs.harvard.edu/abs/2019MNRAS.488.3268A} {488, 3268}

\bibitem[\protect\citeauthoryear{{Beers} \& {Christlieb}}{{Beers} \&
  {Christlieb}}{2005}]{2005ARA&A..43..531B}
{Beers} T.~C.,  {Christlieb} N.,  2005, \mn@doi [\araa]
  {10.1146/annurev.astro.42.053102.134057}, \href
  {https://ui.adsabs.harvard.edu/abs/2005ARA&A..43..531B} {43, 531}

\bibitem[\protect\citeauthoryear{{Behroozi}, {Wechsler}  \& {Wu}}{{Behroozi}
  et~al.}{2013a}]{2013ApJ...762..109B}
{Behroozi} P.~S.,  {Wechsler} R.~H.,   {Wu} H.-Y.,  2013a, \mn@doi [\apj]
  {10.1088/0004-637X/762/2/109}, \href
  {https://ui.adsabs.harvard.edu/abs/2013ApJ...762..109B} {762, 109}

\bibitem[\protect\citeauthoryear{{Behroozi}, {Wechsler}, {Wu}, {Busha},
  {Klypin}  \& {Primack}}{{Behroozi} et~al.}{2013b}]{2013ApJ...763...18B}
{Behroozi} P.~S.,  {Wechsler} R.~H.,  {Wu} H.-Y.,  {Busha} M.~T.,  {Klypin}
  A.~A.,   {Primack} J.~R.,  2013b, \mn@doi [\apj]
  {10.1088/0004-637X/763/1/18}, \href
  {https://ui.adsabs.harvard.edu/abs/2013ApJ...763...18B} {763, 18}

\bibitem[\protect\citeauthoryear{{Berger} \& {Colella}}{{Berger} \&
  {Colella}}{1989}]{1989JCoPh..82...64B}
{Berger} M.~J.,  {Colella} P.,  1989, Journal of Computational Physics, \href
  {http://adsabs.harvard.edu/abs/1989JCoPh..82...64B} {82, 64}

\bibitem[\protect\citeauthoryear{{Bromm} \& {Loeb}}{{Bromm} \&
  {Loeb}}{2003}]{2003Natur.425..812B}
{Bromm} V.,  {Loeb} A.,  2003, \nat, \href
  {http://adsabs.harvard.edu/cgi-bin/nph-bib_query?bibcode=2003Natur.425..812B&db_key=AST}
  {425, 812}

\bibitem[\protect\citeauthoryear{{Brummel-Smith} et~al.,}{{Brummel-Smith}
  et~al.}{2019}]{2019JOSS....4.1636B}
{Brummel-Smith} C.,  et~al., 2019, \mn@doi [The Journal of Open Source
  Software] {10.21105/joss.01636}, \href
  {https://ui.adsabs.harvard.edu/abs/2019JOSS....4.1636B} {4, 1636}

\bibitem[\protect\citeauthoryear{{Bryan} et~al.,}{{Bryan}
  et~al.}{2014}]{2014ApJS..211...19B}
{Bryan} G.~L.,  et~al., 2014, \mn@doi [\apjs] {10.1088/0067-0049/211/2/19},
  \href {https://ui.adsabs.harvard.edu/abs/2014ApJS..211...19B} {211, 19}

\bibitem[\protect\citeauthoryear{{Caffau} et~al.,}{{Caffau}
  et~al.}{2011}]{2011Natur.477...67C}
{Caffau} E.,  et~al., 2011, \mn@doi [\nat] {10.1038/nature10377}, \href
  {https://ui.adsabs.harvard.edu/abs/2011Natur.477...67C} {477, 67}

\bibitem[\protect\citeauthoryear{{Chiaki}, {Marassi}, {Nozawa}, {Yoshida},
  {Schneider}, {Omukai}, {Limongi}  \& {Chieffi}}{{Chiaki}
  et~al.}{2015}]{2015MNRAS.446.2659C}
{Chiaki} G.,  {Marassi} S.,  {Nozawa} T.,  {Yoshida} N.,  {Schneider} R.,
  {Omukai} K.,  {Limongi} M.,   {Chieffi} A.,  2015, \mn@doi [\mnras]
  {10.1093/mnras/stu2298}, \href
  {https://ui.adsabs.harvard.edu/abs/2015MNRAS.446.2659C} {446, 2659}

\bibitem[\protect\citeauthoryear{{Chiaki}, {Susa}  \& {Hirano}}{{Chiaki}
  et~al.}{2018}]{2018MNRAS.475.4378C}
{Chiaki} G.,  {Susa} H.,   {Hirano} S.,  2018, \mn@doi [\mnras]
  {10.1093/mnras/sty040}, \href
  {https://ui.adsabs.harvard.edu/abs/2018MNRAS.475.4378C} {475, 4378}

\bibitem[\protect\citeauthoryear{{Colella} \& {Woodward}}{{Colella} \&
  {Woodward}}{1984}]{1984JCoPh..54..174C}
{Colella} P.,  {Woodward} P.~R.,  1984, Journal of Computational Physics, \href
  {http://adsabs.harvard.edu/abs/1984JCoPh..54..174C} {54, 174}

\bibitem[\protect\citeauthoryear{{Cooke} \& {Madau}}{{Cooke} \&
  {Madau}}{2014}]{2014ApJ...791..116C}
{Cooke} R.~J.,  {Madau} P.,  2014, \mn@doi [\apj]
  {10.1088/0004-637X/791/2/116}, \href
  {https://ui.adsabs.harvard.edu/abs/2014ApJ...791..116C} {791, 116}

\bibitem[\protect\citeauthoryear{{Correa Magnus}, {Smith}, {Khochfar},
  {O'Shea}, {Wise}, {Norman}  \& {Turk}}{{Correa Magnus}
  et~al.}{2024}]{2024MNRAS.527..307C}
{Correa Magnus} L.,  {Smith} B.~D.,  {Khochfar} S.,  {O'Shea} B.~W.,  {Wise}
  J.~H.,  {Norman} M.~L.,   {Turk} M.~J.,  2024, \mn@doi [\mnras]
  {10.1093/mnras/stad3167}, \href
  {https://ui.adsabs.harvard.edu/abs/2024MNRAS.527..307C} {527, 307}

\bibitem[\protect\citeauthoryear{{Crosby}, {O'Shea}, {Smith}, {Turk}  \&
  {Hahn}}{{Crosby} et~al.}{2013}]{2013ApJ...773..108C}
{Crosby} B.~D.,  {O'Shea} B.~W.,  {Smith} B.~D.,  {Turk} M.~J.,   {Hahn} O.,
  2013, \mn@doi [\apj] {10.1088/0004-637X/773/2/108}, \href
  {https://ui.adsabs.harvard.edu/abs/2013ApJ...773..108C} {773, 108}

\bibitem[\protect\citeauthoryear{Dalcin \& Fang}{Dalcin \&
  Fang}{2021}]{mpi4py2}
Dalcin L.,  Fang Y.-L.~L.,  2021, \mn@doi [Computing in Science & Engineering]
  {10.1109/MCSE.2021.3083216}, 23, 47

\bibitem[\protect\citeauthoryear{Dalcin, Paz, Kler  \& Cosimo}{Dalcin
  et~al.}{2011}]{mpi4py3}
Dalcin L.~D.,  Paz R.~R.,  Kler P.~A.,   Cosimo A.,  2011, \mn@doi [Advances in
  Water Resources] {https://doi.org/10.1016/j.advwatres.2011.04.013}, 34, 1124

\bibitem[\protect\citeauthoryear{Dalcín, Paz  \& Storti}{Dalcín
  et~al.}{2005}]{mpi4py5}
Dalcín L.,  Paz R.,   Storti M.,  2005, \mn@doi [Journal of Parallel and
  Distributed Computing] {https://doi.org/10.1016/j.jpdc.2005.03.010}, 65, 1108

\bibitem[\protect\citeauthoryear{Dalcín, Paz, Storti  \& D’Elía}{Dalcín
  et~al.}{2008}]{mpi4py4}
Dalcín L.,  Paz R.,  Storti M.,   D’Elía J.,  2008, \mn@doi [Journal of
  Parallel and Distributed Computing]
  {https://doi.org/10.1016/j.jpdc.2007.09.005}, 68, 655

\bibitem[\protect\citeauthoryear{{Efstathiou}, {Davis}, {White}  \&
  {Frenk}}{{Efstathiou} et~al.}{1985}]{1985ApJS...57..241E}
{Efstathiou} G.,  {Davis} M.,  {White} S.~D.~M.,   {Frenk} C.~S.,  1985,
  \mn@doi [\apjs] {10.1086/191003}, \href
  {http://adsabs.harvard.edu/abs/1985ApJS...57..241E} {57, 241}

\bibitem[\protect\citeauthoryear{{Eisenstein} \& {Hu}}{{Eisenstein} \&
  {Hu}}{1999}]{1999ApJ...511....5E}
{Eisenstein} D.~J.,  {Hu} W.,  1999, \mn@doi [\apj] {10.1086/306640}, \href
  {https://ui.adsabs.harvard.edu/abs/1999ApJ...511....5E} {511, 5}

\bibitem[\protect\citeauthoryear{{Feathers}, {Kulkarni}, {Visbal}  \&
  {Hazlett}}{{Feathers} et~al.}{2024}]{2024ApJ...962...62F}
{Feathers} C.~R.,  {Kulkarni} M.,  {Visbal} E.,   {Hazlett} R.,  2024, \mn@doi
  [\apj] {10.3847/1538-4357/ad1688}, \href
  {https://ui.adsabs.harvard.edu/abs/2024ApJ...962...62F} {962, 62}

\bibitem[\protect\citeauthoryear{{Federrath}, {Sur}, {Schleicher}, {Banerjee}
  \& {Klessen}}{{Federrath} et~al.}{2011}]{2011ApJ...731...62F}
{Federrath} C.,  {Sur} S.,  {Schleicher} D. R.~G.,  {Banerjee} R.,   {Klessen}
  R.~S.,  2011, \mn@doi [\apj] {10.1088/0004-637X/731/1/62}, \href
  {https://ui.adsabs.harvard.edu/abs/2011ApJ...731...62F} {731, 62}

\bibitem[\protect\citeauthoryear{{Ferland} et~al.,}{{Ferland}
  et~al.}{2013}]{2013RMxAA..49..137F}
{Ferland} G.~J.,  et~al., 2013, \mn@doi [\rmxaa] {10.48550/arXiv.1302.4485},
  \href {https://ui.adsabs.harvard.edu/abs/2013RMxAA..49..137F} {49, 137}

\bibitem[\protect\citeauthoryear{{Frebel}, {Johnson}  \& {Bromm}}{{Frebel}
  et~al.}{2007}]{2007MNRAS.380L..40F}
{Frebel} A.,  {Johnson} J.~L.,   {Bromm} V.,  2007, \mn@doi [\mnras]
  {10.1111/j.1745-3933.2007.00344.x}, \href
  {https://ui.adsabs.harvard.edu/abs/2007MNRAS.380L..40F} {380, L40}

\bibitem[\protect\citeauthoryear{{Hahn} \& {Abel}}{{Hahn} \&
  {Abel}}{2011}]{2011MNRAS.415.2101H}
{Hahn} O.,  {Abel} T.,  2011, \mn@doi [\mnras]
  {10.1111/j.1365-2966.2011.18820.x}, \href
  {https://ui.adsabs.harvard.edu/abs/2011MNRAS.415.2101H} {415, 2101}

\bibitem[\protect\citeauthoryear{Harris et~al.,}{Harris
  et~al.}{2020}]{harris2020array}
Harris C.~R.,  et~al., 2020, \mn@doi [Nature] {10.1038/s41586-020-2649-2}, 585,
  357

\bibitem[\protect\citeauthoryear{{Hartwig} \& {Yoshida}}{{Hartwig} \&
  {Yoshida}}{2019}]{2019ApJ...870L...3H}
{Hartwig} T.,  {Yoshida} N.,  2019, \mn@doi [\apjl] {10.3847/2041-8213/aaf866},
  \href {https://ui.adsabs.harvard.edu/abs/2019ApJ...870L...3H} {870, L3}

\bibitem[\protect\citeauthoryear{{Hicks}, {Wells}, {Norman}, {Wise}, {Smith}
  \& {O'Shea}}{{Hicks} et~al.}{2021}]{2021ApJ...909...70H}
{Hicks} W.~M.,  {Wells} A.,  {Norman} M.~L.,  {Wise} J.~H.,  {Smith} B.~D.,
  {O'Shea} B.~W.,  2021, \mn@doi [\apj] {10.3847/1538-4357/abda3a}, \href
  {https://ui.adsabs.harvard.edu/abs/2021ApJ...909...70H} {909, 70}

\bibitem[\protect\citeauthoryear{{Hirano}, {Hosokawa}, {Yoshida}, {Umeda},
  {Omukai}, {Chiaki}  \& {Yorke}}{{Hirano} et~al.}{2014}]{2014ApJ...781...60H}
{Hirano} S.,  {Hosokawa} T.,  {Yoshida} N.,  {Umeda} H.,  {Omukai} K.,
  {Chiaki} G.,   {Yorke} H.~W.,  2014, \mn@doi [\apj]
  {10.1088/0004-637X/781/2/60}, \href
  {https://ui.adsabs.harvard.edu/abs/2014ApJ...781...60H} {781, 60}

\bibitem[\protect\citeauthoryear{{Hirashita} \& {Ferrara}}{{Hirashita} \&
  {Ferrara}}{2005}]{2005MNRAS.356.1529H}
{Hirashita} H.,  {Ferrara} A.,  2005, \mn@doi [\mnras]
  {10.1111/j.1365-2966.2004.08602.x}, \href
  {https://ui.adsabs.harvard.edu/abs/2005MNRAS.356.1529H} {356, 1529}

\bibitem[\protect\citeauthoryear{{Hockney} \& {Eastwood}}{{Hockney} \&
  {Eastwood}}{1988}]{1988csup.book.....H}
{Hockney} R.~W.,  {Eastwood} J.~W.,  1988, {Computer simulation using
  particles}.
{CRC Press}

\bibitem[\protect\citeauthoryear{{Hollenbach} \& {McKee}}{{Hollenbach} \&
  {McKee}}{1979}]{1979ApJS...41..555H}
{Hollenbach} D.,  {McKee} C.~F.,  1979, \mn@doi [\apjs] {10.1086/190631}, \href
  {https://ui.adsabs.harvard.edu/abs/1979ApJS...41..555H} {41, 555}

\bibitem[\protect\citeauthoryear{{Hollenbach} \& {McKee}}{{Hollenbach} \&
  {McKee}}{1989}]{1989ApJ...342..306H}
{Hollenbach} D.,  {McKee} C.~F.,  1989, \mn@doi [\apj] {10.1086/167595}, \href
  {https://ui.adsabs.harvard.edu/abs/1989ApJ...342..306H} {342, 306}

\bibitem[\protect\citeauthoryear{{Holzbauer} \& {Furlanetto}}{{Holzbauer} \&
  {Furlanetto}}{2012}]{2012MNRAS.419..718H}
{Holzbauer} L.~N.,  {Furlanetto} S.~R.,  2012, \mn@doi [\mnras]
  {10.1111/j.1365-2966.2011.19752.x}, \href
  {https://ui.adsabs.harvard.edu/abs/2012MNRAS.419..718H} {419, 718}

\bibitem[\protect\citeauthoryear{Hunter}{Hunter}{2007}]{Hunter2007}
Hunter J.~D.,  2007, \mn@doi [Computing in Science \& Engineering]
  {10.1109/MCSE.2007.55}, 9, 90

\bibitem[\protect\citeauthoryear{{Incatasciato}, {Khochfar}  \&
  {O{\~n}orbe}}{{Incatasciato} et~al.}{2023}]{2023MNRAS.522..330I}
{Incatasciato} A.,  {Khochfar} S.,   {O{\~n}orbe} J.,  2023, \mn@doi [\mnras]
  {10.1093/mnras/stad1008}, \href
  {https://ui.adsabs.harvard.edu/abs/2023MNRAS.522..330I} {522, 330}

\bibitem[\protect\citeauthoryear{{Ishigaki}, {Tominaga}, {Kobayashi}  \&
  {Nomoto}}{{Ishigaki} et~al.}{2014}]{2014ApJ...792L..32I}
{Ishigaki} M.~N.,  {Tominaga} N.,  {Kobayashi} C.,   {Nomoto} K.,  2014,
  \mn@doi [\apjl] {10.1088/2041-8205/792/2/L32}, \href
  {https://ui.adsabs.harvard.edu/abs/2014ApJ...792L..32I} {792, L32}

\bibitem[\protect\citeauthoryear{{Jeena}, {Banerjee}, {Chiaki}  \&
  {Heger}}{{Jeena} et~al.}{2023}]{2023MNRAS.526.4467J}
{Jeena} S.~K.,  {Banerjee} P.,  {Chiaki} G.,   {Heger} A.,  2023, \mn@doi
  [\mnras] {10.1093/mnras/stad3028}, \href
  {https://ui.adsabs.harvard.edu/abs/2023MNRAS.526.4467J} {526, 4467}

\bibitem[\protect\citeauthoryear{{Ji}, {Frebel}  \& {Bromm}}{{Ji}
  et~al.}{2015}]{2015MNRAS.454..659J}
{Ji} A.~P.,  {Frebel} A.,   {Bromm} V.,  2015, \mn@doi [\mnras]
  {10.1093/mnras/stv2052}, \href
  {https://ui.adsabs.harvard.edu/abs/2015MNRAS.454..659J} {454, 659}

\bibitem[\protect\citeauthoryear{{Johnson}, {Dalla Vecchia}  \&
  {Khochfar}}{{Johnson} et~al.}{2013}]{2013MNRAS.428.1857J}
{Johnson} J.~L.,  {Dalla Vecchia} C.,   {Khochfar} S.,  2013, \mn@doi [\mnras]
  {10.1093/mnras/sts011}, \href
  {https://ui.adsabs.harvard.edu/abs/2013MNRAS.428.1857J} {428, 1857}

\bibitem[\protect\citeauthoryear{{Klessen} \& {Glover}}{{Klessen} \&
  {Glover}}{2023}]{2023ARA&A..61...65K}
{Klessen} R.~S.,  {Glover} S. C.~O.,  2023, \mn@doi [\araa]
  {10.1146/annurev-astro-071221-053453}, \href
  {https://ui.adsabs.harvard.edu/abs/2023ARA&A..61...65K} {61, 65}

\bibitem[\protect\citeauthoryear{{Komatsu} et~al.,}{{Komatsu}
  et~al.}{2011}]{2011ApJS..192...18K}
{Komatsu} E.,  et~al., 2011, \mn@doi [\apjs] {10.1088/0067-0049/192/2/18},
  \href {https://ui.adsabs.harvard.edu/abs/2011ApJS..192...18K} {192, 18}

\bibitem[\protect\citeauthoryear{{Krumholz}}{{Krumholz}}{2014a}]{2014MNRAS.437.1662K}
{Krumholz} M.~R.,  2014a, \mn@doi [\mnras] {10.1093/mnras/stt2000}, \href
  {https://ui.adsabs.harvard.edu/abs/2014MNRAS.437.1662K} {437, 1662}

\bibitem[\protect\citeauthoryear{{Krumholz}}{{Krumholz}}{2014b}]{2014PhR...539...49K}
{Krumholz} M.~R.,  2014b, \mn@doi [\physrep] {10.1016/j.physrep.2014.02.001},
  \href {https://ui.adsabs.harvard.edu/abs/2014PhR...539...49K} {539, 49}

\bibitem[\protect\citeauthoryear{{Kulkarni}, {Visbal}  \& {Bryan}}{{Kulkarni}
  et~al.}{2021}]{2021ApJ...917...40K}
{Kulkarni} M.,  {Visbal} E.,   {Bryan} G.~L.,  2021, \mn@doi [\apj]
  {10.3847/1538-4357/ac08a3}, \href
  {https://ui.adsabs.harvard.edu/abs/2021ApJ...917...40K} {917, 40}

\bibitem[\protect\citeauthoryear{{Lau}, {Kravtsov}  \& {Nagai}}{{Lau}
  et~al.}{2009}]{2009ApJ...705.1129L}
{Lau} E.~T.,  {Kravtsov} A.~V.,   {Nagai} D.,  2009, \mn@doi [\apj]
  {10.1088/0004-637X/705/2/1129}, \href
  {https://ui.adsabs.harvard.edu/abs/2009ApJ...705.1129L} {705, 1129}

\bibitem[\protect\citeauthoryear{{Maio}, {Khochfar}, {Johnson}  \&
  {Ciardi}}{{Maio} et~al.}{2011}]{2011MNRAS.414.1145M}
{Maio} U.,  {Khochfar} S.,  {Johnson} J.~L.,   {Ciardi} B.,  2011, \mn@doi
  [\mnras] {10.1111/j.1365-2966.2011.18455.x}, \href
  {https://ui.adsabs.harvard.edu/abs/2011MNRAS.414.1145M} {414, 1145}

\bibitem[\protect\citeauthoryear{{Meece}, {Smith}  \& {O'Shea}}{{Meece}
  et~al.}{2014}]{2014ApJ...783...75M}
{Meece} G.~R.,  {Smith} B.~D.,   {O'Shea} B.~W.,  2014, \mn@doi [\apj]
  {10.1088/0004-637X/783/2/75}, \href
  {https://ui.adsabs.harvard.edu/abs/2014ApJ...783...75M} {783, 75}

\bibitem[\protect\citeauthoryear{{Nomoto}, {Tominaga}, {Umeda}, {Kobayashi}  \&
  {Maeda}}{{Nomoto} et~al.}{2006}]{2006NuPhA.777..424N}
{Nomoto} K.,  {Tominaga} N.,  {Umeda} H.,  {Kobayashi} C.,   {Maeda} K.,  2006,
  \mn@doi [\nphysa] {10.1016/j.nuclphysa.2006.05.008}, \href
  {https://ui.adsabs.harvard.edu/abs/2006NuPhA.777..424N} {777, 424}

\bibitem[\protect\citeauthoryear{{Oh} \& {Haiman}}{{Oh} \&
  {Haiman}}{2002}]{2002ApJ...569..558O}
{Oh} S.~P.,  {Haiman} Z.,  2002, \mn@doi [\apj] {10.1086/339393}, \href
  {https://ui.adsabs.harvard.edu/abs/2002ApJ...569..558O} {569, 558}

\bibitem[\protect\citeauthoryear{{Omukai}}{{Omukai}}{2000}]{2000ApJ...534..809O}
{Omukai} K.,  2000, \mn@doi [\apj] {10.1086/308776}, \href
  {http://adsabs.harvard.edu/cgi-bin/nph-bib_query?bibcode=2000ApJ...534..809O&db_key=AST}
  {534, 809}

\bibitem[\protect\citeauthoryear{{Omukai}, {Tsuribe}, {Schneider}  \&
  {Ferrara}}{{Omukai} et~al.}{2005}]{2005ApJ...626..627O}
{Omukai} K.,  {Tsuribe} T.,  {Schneider} R.,   {Ferrara} A.,  2005, \mn@doi
  [\apj] {10.1086/429955}, \href
  {http://adsabs.harvard.edu/cgi-bin/nph-bib_query?bibcode=2005ApJ...626..627O&db_key=AST}
  {626, 627}

\bibitem[\protect\citeauthoryear{{Pollack}, {Hollenbach}, {Beckwith},
  {Simonelli}, {Roush}  \& {Fong}}{{Pollack}
  et~al.}{1994}]{1994ApJ...421..615P}
{Pollack} J.~B.,  {Hollenbach} D.,  {Beckwith} S.,  {Simonelli} D.~P.,  {Roush}
  T.,   {Fong} W.,  1994, \mn@doi [\apj] {10.1086/173677}, \href
  {https://ui.adsabs.harvard.edu/abs/1994ApJ...421..615P} {421, 615}

\bibitem[\protect\citeauthoryear{Rogowski, Aseeri, Keyes  \& Dalcin}{Rogowski
  et~al.}{2023}]{mpi4py1}
Rogowski M.,  Aseeri S.,  Keyes D.,   Dalcin L.,  2023, \mn@doi [IEEE
  Transactions on Parallel and Distributed Systems]
  {10.1109/TPDS.2022.3225481}, 34, 611

\bibitem[\protect\citeauthoryear{{Schaerer}}{{Schaerer}}{2002}]{2002A&A...382...28S}
{Schaerer} D.,  2002, \mn@doi [\aap] {10.1051/0004-6361:20011619}, \href
  {https://ui.adsabs.harvard.edu/abs/2002A&A...382...28S} {382, 28}

\bibitem[\protect\citeauthoryear{{Schauer}, {Glover}, {Klessen}  \&
  {Clark}}{{Schauer} et~al.}{2021}]{2021MNRAS.507.1775S}
{Schauer} A. T.~P.,  {Glover} S. C.~O.,  {Klessen} R.~S.,   {Clark} P.,  2021,
  \mn@doi [\mnras] {10.1093/mnras/stab1953}, \href
  {https://ui.adsabs.harvard.edu/abs/2021MNRAS.507.1775S} {507, 1775}

\bibitem[\protect\citeauthoryear{{Schneider}, {Omukai}, {Bianchi}  \&
  {Valiante}}{{Schneider} et~al.}{2012}]{2012MNRAS.419.1566S}
{Schneider} R.,  {Omukai} K.,  {Bianchi} S.,   {Valiante} R.,  2012, \mn@doi
  [\mnras] {10.1111/j.1365-2966.2011.19818.x}, \href
  {https://ui.adsabs.harvard.edu/abs/2012MNRAS.419.1566S} {419, 1566}

\bibitem[\protect\citeauthoryear{{Sharda} \& {Krumholz}}{{Sharda} \&
  {Krumholz}}{2022}]{2022MNRAS.509.1959S}
{Sharda} P.,  {Krumholz} M.~R.,  2022, \mn@doi [\mnras]
  {10.1093/mnras/stab2921}, \href
  {https://ui.adsabs.harvard.edu/abs/2022MNRAS.509.1959S} {509, 1959}

\bibitem[\protect\citeauthoryear{{Sluder}, {Ritter}, {Safranek-Shrader},
  {Milosavljevi{\'c}}  \& {Bromm}}{{Sluder} et~al.}{2016}]{2016MNRAS.456.1410S}
{Sluder} A.,  {Ritter} J.~S.,  {Safranek-Shrader} C.,  {Milosavljevi{\'c}} M.,
   {Bromm} V.,  2016, \mn@doi [\mnras] {10.1093/mnras/stv2587}, \href
  {https://ui.adsabs.harvard.edu/abs/2016MNRAS.456.1410S} {456, 1410}

\bibitem[\protect\citeauthoryear{Smith \& Lang}{Smith \& Lang}{2019}]{ytree}
Smith B.~D.,  Lang M.,  2019, \mn@doi [Journal of Open Source Software]
  {10.21105/joss.01881}, 4, 1881

\bibitem[\protect\citeauthoryear{{Smith}, {Sigurdsson}  \& {Abel}}{{Smith}
  et~al.}{2008}]{2008MNRAS.385.1443S}
{Smith} B.,  {Sigurdsson} S.,   {Abel} T.,  2008, \mn@doi [\mnras]
  {10.1111/j.1365-2966.2008.12922.x}, \href
  {http://adsabs.harvard.edu/abs/2008MNRAS.385.1443S} {385, 1443}

\bibitem[\protect\citeauthoryear{{Smith}, {Wise}, {O'Shea}, {Norman}  \&
  {Khochfar}}{{Smith} et~al.}{2015}]{2015MNRAS.452.2822S}
{Smith} B.~D.,  {Wise} J.~H.,  {O'Shea} B.~W.,  {Norman} M.~L.,   {Khochfar}
  S.,  2015, \mn@doi [\mnras] {10.1093/mnras/stv1509}, \href
  {https://ui.adsabs.harvard.edu/abs/2015MNRAS.452.2822S} {452, 2822}

\bibitem[\protect\citeauthoryear{{Smith} et~al.,}{{Smith}
  et~al.}{2017}]{2017MNRAS.466.2217S}
{Smith} B.~D.,  et~al., 2017, \mn@doi [\mnras] {10.1093/mnras/stw3291}, \href
  {https://ui.adsabs.harvard.edu/abs/2017MNRAS.466.2217S} {466, 2217}

\bibitem[\protect\citeauthoryear{Smith et~al.,}{Smith et~al.}{2023a}]{yt-astro}
Smith B.,  et~al., 2023a, {yt-project/yt\_astro\_analysis:
  yt\_astro\_analysis-1.1.3}, \mn@doi{10.5281/zenodo.8431185}, \url
  {https://doi.org/10.5281/zenodo.8431185}

\bibitem[\protect\citeauthoryear{Smith, Lang, Wise  \& Bazán}{Smith
  et~al.}{2023b}]{ytree-zenodo}
Smith B.,  Lang M.,  Wise J.,   Bazán J.,  2023b, ytree-project/ytree: ytree
  3.2.1 Release, \mn@doi{10.5281/zenodo.8349044}, \url
  {https://doi.org/10.5281/zenodo.8349044}

\bibitem[\protect\citeauthoryear{{Starkenburg} et~al.,}{{Starkenburg}
  et~al.}{2018}]{2018MNRAS.481.3838S}
{Starkenburg} E.,  et~al., 2018, \mn@doi [\mnras] {10.1093/mnras/sty2276},
  \href {https://ui.adsabs.harvard.edu/abs/2018MNRAS.481.3838S} {481, 3838}

\bibitem[\protect\citeauthoryear{{Turk}, {Smith}, {Oishi}, {Skory}, {Skillman},
  {Abel}  \& {Norman}}{{Turk} et~al.}{2011}]{2011ApJS..192....9T}
{Turk} M.~J.,  {Smith} B.~D.,  {Oishi} J.~S.,  {Skory} S.,  {Skillman} S.~W.,
  {Abel} T.,   {Norman} M.~L.,  2011, \mn@doi [\apjs]
  {10.1088/0067-0049/192/1/9}, \href
  {https://ui.adsabs.harvard.edu/abs/2011ApJS..192....9T} {192, 9}

\bibitem[\protect\citeauthoryear{Van~Rossum \& Drake~Jr}{Van~Rossum \&
  Drake~Jr}{1995}]{van1995python}
Van~Rossum G.,  Drake~Jr F.~L.,  1995, Python reference manual.
Centrum voor Wiskunde en Informatica Amsterdam

\bibitem[\protect\citeauthoryear{{Voit}, {Donahue}, {Bryan}  \&
  {McDonald}}{{Voit} et~al.}{2015}]{2015Natur.519..203V}
{Voit} G.~M.,  {Donahue} M.,  {Bryan} G.~L.,   {McDonald} M.,  2015, \mn@doi
  [\nat] {10.1038/nature14167}, \href
  {https://ui.adsabs.harvard.edu/abs/2015Natur.519..203V} {519, 203}

\bibitem[\protect\citeauthoryear{{Wise} \& {Abel}}{{Wise} \&
  {Abel}}{2011}]{2011MNRAS.414.3458W}
{Wise} J.~H.,  {Abel} T.,  2011, \mn@doi [\mnras]
  {10.1111/j.1365-2966.2011.18646.x}, \href
  {https://ui.adsabs.harvard.edu/abs/2011MNRAS.414.3458W} {414, 3458}

\bibitem[\protect\citeauthoryear{{Wise}, {Turk}, {Norman}  \& {Abel}}{{Wise}
  et~al.}{2012}]{2012ApJ...745...50W}
{Wise} J.~H.,  {Turk} M.~J.,  {Norman} M.~L.,   {Abel} T.,  2012, \mn@doi
  [\apj] {10.1088/0004-637X/745/1/50}, \href
  {https://ui.adsabs.harvard.edu/abs/2012ApJ...745...50W} {745, 50}

\bibitem[\protect\citeauthoryear{{Wise}, {Demchenko}, {Halicek}, {Norman},
  {Turk}, {Abel}  \& {Smith}}{{Wise} et~al.}{2014}]{2014MNRAS.442.2560W}
{Wise} J.~H.,  {Demchenko} V.~G.,  {Halicek} M.~T.,  {Norman} M.~L.,  {Turk}
  M.~J.,  {Abel} T.,   {Smith} B.~D.,  2014, \mn@doi [\mnras]
  {10.1093/mnras/stu979}, \href
  {https://ui.adsabs.harvard.edu/abs/2014MNRAS.442.2560W} {442, 2560}

\bibitem[\protect\citeauthoryear{{Wolcott-Green}, {Haiman}  \&
  {Bryan}}{{Wolcott-Green} et~al.}{2011}]{2011MNRAS.418..838W}
{Wolcott-Green} J.,  {Haiman} Z.,   {Bryan} G.~L.,  2011, \mn@doi [\mnras]
  {10.1111/j.1365-2966.2011.19538.x}, \href
  {https://ui.adsabs.harvard.edu/abs/2011MNRAS.418..838W} {418, 838}

\bibitem[\protect\citeauthoryear{{de Bennassuti}, {Salvadori}, {Schneider},
  {Valiante}  \& {Omukai}}{{de Bennassuti} et~al.}{2017}]{2017MNRAS.465..926D}
{de Bennassuti} M.,  {Salvadori} S.,  {Schneider} R.,  {Valiante} R.,
  {Omukai} K.,  2017, \mn@doi [\mnras] {10.1093/mnras/stw2687}, \href
  {https://ui.adsabs.harvard.edu/abs/2017MNRAS.465..926D} {465, 926}

\makeatother
\end{thebibliography}

\bsp	
\label{lastpage}
\end{document}